\documentclass[11pt,a4paper]{article}
\usepackage[english]{babel}
\usepackage[utf8]{inputenc}
\usepackage{url}
\usepackage{epsf}
\usepackage{cite}
\usepackage{graphicx}
\usepackage{subcaption} 

\usepackage{amsmath}
\usepackage{amssymb}
\usepackage{mathrsfs}
\usepackage[normalem]{ulem}
\usepackage{mathbbol}
\usepackage{multirow}
\usepackage{slashed}
\usepackage{amsfonts}
\usepackage{bbding}
\usepackage{array}
\usepackage[
colorlinks=true
,urlcolor=black
,anchorcolor=black
,citecolor=black
,filecolor=black
,linkcolor=black
,menucolor=black
,linktocpage=true
,pdfproducer=medialab
,pdfa=true
]{hyperref}

\usepackage{color} 
\usepackage[left=2cm,right=2cm,top=2cm,bottom=2cm]{geometry}
\usepackage{tikz}
\usetikzlibrary{shapes,arrows,positioning,automata,backgrounds,calc,er,patterns}
\usepackage[compat=1.1.0]{tikz-feynman}
\usepackage{contour}
\allowdisplaybreaks
\begin{document}
\begin{center}
\vspace*{1mm}
\vspace{1.3cm}
{\Large\bf
High-energy cLFV at $\mathbf{\mu}$TRISTAN: }

\vspace{3mm}
{\Large\bf
 HNL extensions of the Standard Model
}

\vspace*{1.2cm}

{\bf J.~Kriewald $^\text{a}$, E.~Pinsard $^\text{b}$ and A.~M.~Teixeira $^\text{c}$ }

\vspace*{.5cm}
$^\text{a}$ {Jožef Stefan Institut, Jamova Cesta 39, P. O. Box 3000, 1001 Ljubljana, Slovenia}

\vspace*{.2cm}
$^\text{b}$ {Physik-Institut, Universität Zürich, CH-8057 Zürich, Switzerland}

\vspace*{.2cm}
$^\text{c}$ Laboratoire de Physique de Clermont Auvergne (UMR 6533), CNRS/IN2P3,\\
Univ. Clermont Auvergne, 4 Av. Blaise Pascal, 63178 Aubi\`ere Cedex,
France

\end{center}

\vspace*{5mm}
\begin{abstract}
\noindent
Within the context of heavy neutral lepton (HNL) extensions of the Standard Model,
we 
compute the cross-sections for $\mu^+ e^-\to \ell_\alpha^+\ell_\beta^-$ scattering, as well as several angular observables.
In particular, we investigate the future sensitivity of a $\mu$TRISTAN collider in discovering such charged lepton flavour violating processes and the potential constraining power of these searches on the parameter space of HNL models.
Our results show that while low-energy probes of $\mu-e$ flavour violation do offer the most promising potential,  
the prospects for $e\tau$ and $\mu\tau$ flavour violation searches at $\mu$TRISTAN can exceed those of related low-energy probes (as well as flavour violating $Z$-pole processes at FCC-ee) by several orders of magnitude.
\end{abstract}

\section{Introduction}
Heavy neutral leptons (HNL) are among the most appealing and yet most simple extensions of the Standard Model (SM). In particular, HNL (or heavy sterile fermions) are present in numerous extensions of the SM aiming at explaining neutrino mass generation: this is the case of minimal seesaw models, such as the type I~\cite{Minkowski:1977sc,Yanagida:1979as,Glashow:1979nm,Gell-Mann:1979vob,Mohapatra:1979ia}, or variations of the latter, including realisations of the Linear Seesaw~\cite{Akhmedov:1995vm, Barr:2003nn,Malinsky:2005bi}, or of the Inverse Seesaw~\cite{Schechter:1980gr, Gronau:1984ct, Mohapatra:1986bd}. These heavy sterile states also emerge in the context of ambitious New Physics (NP) constructions, as for instance Left-Right symmetric extensions of the SM, or in Grand Unified Theories. 
Furthermore, their role in cosmology has also been intensively explored, as they might also play a role in explaining the observed baryon asymmetry of the Universe (via leptogenesis), or possibly account for (part) of the dark matter relic abundance. 

In view of the manifest physics potential of these states, dedicated searches are being currently carried out, directly at the LHC, or indirectly via their role on a large number of high-intensity and/or electroweak precision observables. 
The physics cases and programmes of future colliders also prominently feature searches for these states. 

Should the HNLs exhibit non-negligible couplings to the (light) active neutrinos, then significant contributions to charged lepton flavour violating (cLFV) decays and transitions can be expected:
the role of HNLs in low-energy cLFV processes has been extensively studied, be it in the framework of complete models, or then 
in minimal ad-hoc constructions (in which the presence of HNL is encoded in an enlarged spectrum and in new mixings to the active neutrinos)~\cite{Riemann:1982rq,Illana:1999ww,Mann:1983dv,Illana:2000ic,Alonso:2012ji,Ilakovac:1994kj,Ma:1979px,Gronau:1984ct,Deppisch:2004fa,Deppisch:2005zm,Dinh:2012bp,Hambye:2013jsa,Abada:2014kba,Abada:2015oba,Abada:2015zea,Abada:2016vzu,Calibbi:2017uvl,Abada:2018nio,Arganda:2014dta,Marcano:2019rmk,Calderon:2022alb}. 
Searches for heavy sterile fermions at the high-energy frontier have also been intensively pursued in recent years: the physics programme of possible future colliders such as FCC-ee features numerous study cases, see e.g.~\cite{ Blondel:2014bra,Antusch:2016ejd,Cai:2017mow,Blondel:2022qqo,Abdullahi:2022jlv,Abada:2022wvh,Giffin:2022rei,Drewes:2022rsk,Ovchynnikov:2023wgg,Ajmal:2024kwi,Antusch:2024otj} and references therein; likewise, searches for HNLs at a future Muon Collider have also been under recent scrutiny~\cite{Chakraborty:2022pcc,Mekala:2023diu,Kwok:2023dck,Li:2023tbx,Mikulenko:2023ezx,Urquia-Calderon:2023dkf,deLima:2024ohf,Marcos:2024yfm}. 

The case for an asymmetric electron-muon collider has also gained renewed momentum, in view of its potential as a uniquely powerful tool 
for cLFV searches, as well as precise measurements in the Higgs sector. The asymmetric nature of the colliding beams substantially reduces (physics) backgrounds  compared to both electron-positron or muon-antimuon collisions: for $e^- \mu^+$, SM-like final states mostly arise from higher order processes (including vector boson scattering and fusion). Such a facility - $\mu$TRISTAN - has been proposed with several operating collision energies, possibly paving the way for a muon collider~\cite{Hamada:2022mua,Lu:2020dkx}. As already discussed~\cite{Bossi:2020yne}, an asymmetric collider would naturally offer the perfect testing grounds for (tree-level) flavour-violating new physics mediators, as is for instance the case of $Z^\prime$ models~\cite{Kriewald:2022erk,Goudelis:2023yni} or flavour violating neutral and doubly charged scalars~\cite{Dev:2023nha}. 
Likewise, the possibility of a same-sign lepton collider (e.g. $\mu^+\mu^+$) allows to resonantly probe lepton number violation (LNV) by two units (for example in processes such as $\mu^+\mu^+\to W^+W^+\to jjjj$ which can arise in different SM extensions encompassing a mechanism of Majorana neutrino mass generation).
Singly or doubly resonant behaviour in $m_{jj}$ or even $m_{jjjj}$ then allows to distinguish different mediators and topologies, such as a doubly charged scalar in the s-channel or a heavy Majorana fermion mediator in a t-channel (see e.g.~\cite{Dev:2023nha}). 
(Pseudo-) Observables constructed from event-shape variables of such scatterings offer probes of the neutrino nature, which are highly complementary to neutrinoless double beta decay (similar observables at hadron colliders have been recently studied in~\cite{2408.00833}).

In the framework of a minimal type I seesaw, a recent study has focussed on final states (signal and SM background) including charged leptons, missing energy (light neutrinos) and jets~\cite{2410.21956}; limits were inferred 
on active-sterile neutrino mixing angles, leading to new constraints potentially far stronger 
(two orders of magnitude) than those offered by studies of electroweak precision observables (EWPO).

In the present study we focus on the potential of an asymmetric electron-muon collider to search for heavy neutral leptons via (almost) background-free cLFV final state signatures. Albeit generating cLFV only at the 1-loop level,
HNL with non-negligible couplings to active neutrinos can lead to a sizeable 
number of signal events. 
Interestingly, the underlying topologies offer a direct connection to low-energy cLFV observables, and/or to flavour violating $Z \to \ell_\alpha \ell_\beta$ (with $\alpha \neq \beta$) decays. 
Moreover, the asymmetric nature of the beams might be instrumental in probing the nature of the interaction at the source of the cLFV transition. 
In what follows, we thus consider all relevant topologies issuing from 
$\mu^+e^-  \to \ell_\alpha^+ \ell_\beta^-$ scatterings, including s- and t-channel penguin processes, as well as box diagrams. 
We offer a thorough description of the computation, and provide a number of useful expressions for future dedicated studies.  
In addition to the scattering cross-sections (leading to a very conservative estimation of the number of expected events), we also consider pseudo-rapidity distributions, and the forward-backward asymmetry,  
which - as we will subsequently discuss - offer additional insight onto the nature of the underlying processes.

Complementary to the $\mu$TRISTAN observables mentioned above, we also consider a large number of low-energy cLFV transitions and decays (leptonic radiative and three-body decays, muon-electron conversion in nuclei and muonium oscillations) as well as cLFV $Z$ boson decays,   which can be searched for at the optimal environment of an FCC-ee in its $Z$-pole runs. 

In order to illustrate the probing potential of $\mu$TRISTAN in (indirect) cLFV searches for heavy neutral leptons, we consider a simplified NP model, in which two sterile states are added (in an ad-hoc manner) to the SM content; no assumption is made on the mechanism at the origin of the massive neutral lepton spectrum, the effect of the new heavy states being encoded in a generalised leptonic mixing matrix which extends the usual Pontecorvo–Maki–Nakagawa–Sakata (PMNS) matrix, $U_\text{PMNS}$. For completeness, we also study a UV-complete model of neutrino mass generation, which can be realised at low-energies without incurring in a loss of naturalness (in the sense of 't Hooft) - the Inverse Seesaw mechanism, specifically its (3,3) realisation, in which three generation of distinct species of HNL are added to the SM content. 
For both scenarios, we consider illustrative benchmark scenarios for spectrum and flavour patterns.
In what concerns the nominal operating $\mu$TRISTAN beam energies, several possibilities will be explored,  leading to centre-of-mass (c.o.m.) energies from 100~GeV to 1100~GeV.

As we proceed to discuss, even for very conservative estimates of $\mu$TRISTAN efficiency, one can expect a sizeable number of signal events while suppressing possible SM backgrounds to negligible rates: in particular, the t-channel dominated $\mu^+e^-\to \tau^+ e^-$ and $\mu^+e^-\to \mu^+ \tau^-$ scattering processes lead to tens or hundreds of thousands of events in large regions of the parameter space.
We further point out that, should a $\mu^+e^-\to \ell_\alpha^+\ell_\alpha^-$ scattering process be discovered, measurements of the forward-backward asymmetry of the final state leptons would grant access to further constrain the mass scale of the HNL via the interference of s- and t-channel contributions. 
(Notice that this interference behaviour can further be altered by the presence of additional CP-violating phases of the enlarged lepton mixing matrix.)

Finally, we analytically calculate the pseudo-rapidity distributions of the scattering cross-sections allowing to implement cuts in a straightforward manner.
Based on these observations we suggest a basic and conservative ``cut-and-count'' analysis which allows us to estimate future sensitivities of a detector with ATLAS-like rapidity coverage to discover (or rather place bounds on) the high-energy cLFV observables.

In general, our findings suggest that while dedicated low-energy searches for $\mu-e$ 
flavour violation - in particular $\mu-e$ conversion in Nuclei - unquestionably offer the most promising probes for HNL at the origin of cLFV transitions,  
$\mu$TRISTAN has a clear  potential to outperform dedicated (low-energy) searches for $e\tau$ and $\mu\tau$ flavour violation - including $Z$-pole searches at FCC-ee.

The manuscript is organised as follows: we begin by a brief description of $\mu$TRISTAN, focusing on its c.o.m. energies and expected luminosity. 
Section~\ref{sec:muTobservables} is devoted to a detailed computation of high-energy observables which can be studied at $\mu$TRISTAN. 
A brief description of the SM extensions via heavy sterile fermions here considered is offered in Section~\ref{sec:HNL:model:constraints}. Our results and discussion are presented in Section~\ref{sec:results}, followed by final remarks and prospective discussion. 
Additional (detailed and pedagogical) information outlining the calculation of the amplitudes is offered in several appendices. 
\section{An asymmetric lepton collider: $\mu$TRISTAN}\label{sec:muT}
Recently, muon colliders have been the focus of intense discussion, as they would allow reaching higher energies than electron-positron colliders, offering also a much reduced expected background in comparison to $pp$ colliders; if technically feasible, muon colliders could provide a perfect environment for successful direct searches for heavy new resonances and for impressive precision measurements.
However, while the methods for achieving a focused and coherent $\mu^-$ beam 
are still under investigation, the technology to create low-emittance $\mu^+$ beam, using ultra cold muons~\cite{Kondo:2018rzx} is already established.
Thus, relying on the technology developed for the muon $g-2$ experiment at J-PARC~\cite{Abe:2019thb}, one can envisage a $\mu^+\mu^+$ as well as an asymmetric $\mu^+e^-$ collider - the two proposed configurations for $\mu$TRISTAN. In the following we only focus on the asymmetric $\mu^+e^-$ mode, since the $\mu^+\mu^+$ configuration would have a much lower luminosity. 

Assuming realistic operating parameters (see~\cite{Hamada:2022mua} for a 3~km ring design), one can expect that the instantaneous luminosity for the collisions, i.e.
\begin{equation}\label{eq:instLuminosity}
\mathcal{L}^\text{inst}_{\mu^+ e^-}\, =\, 
\frac{N_{\mu^+}\, N_{e^-}}{2\pi \, 
\sqrt{\sigma^2_{\mu^+ x} + \sigma^2_{e^- x} }\, \sqrt{\sigma^2_{\mu^+ y} + \sigma^2_{e^- y} } } \times f_\text{rep}\,,
\end{equation}
in which $f_\text{rep}$ denotes the collision frequency, $N_{\ell}$ the number of leptons per bunch and 
$\sigma^2_{\ell+ }$ the beams' dimensions, might allow achieving $\mathcal{L}^\text{inst}_{\mu^+ e^-} \sim \mathcal{O}(10^{34})\text{cm}^{-2}\text{s}^{-1}$~\cite{1306.6353,FCC:2018evy,1809.00285}, or even higher, for associated centre-of-mass energies ranging from a few hundred to a few thousand~GeV. 
Nominal $(E_{e^-}, E_{\mu^+})$ beam energies have been proposed,  for example 
(20~GeV, 200~GeV), (30~GeV, 1~TeV),  (50~GeV, 1~TeV) or (100~GeV, 3~TeV)~\cite{Hamada:2022mua,Lu:2020dkx,2210.11083} (see Table~\ref{tab:muT:com}), which we will subsequently explore in our phenomenological discussion.

\renewcommand{\arraystretch}{1.3}
\begin{table}[h!]
    \centering
    \begin{tabular}{|c|c|c|}
    \hline
        $E_{e^-}$ (GeV) & $E_{\mu^+}$ (GeV)&  $\sqrt{s}$ (GeV)
         \\ \hline \hline
          20& 200 & 126.5  \\ \hline
          30 & 1000 & 346.4  \\ \hline
            50 & 1000 & 447.2  \\ \hline
           100 & 3000 & 1095.4 \\ \hline
    \end{tabular}
    \caption{Different $e^-$ and $\mu^+$ beam energies considered, as well as the associated centre-of-mass energies.}
    \label{tab:muT:com}
\end{table}
\renewcommand{\arraystretch}{1.}
Even in the absence of a full knowledge of the accelerator characteristics, 
a realistic estimate of the instantaneous luminosity for $\mu$TRISTAN in its (30 GeV, 1 TeV) configuration has been given in~\cite{Hamada:2022mua};
in what follows we rely on their result - $\mathcal{L}_\mathrm{inst} = 4.6 \times 10^{33}$ cm$^{-2}$ s$^{-1}$ - thus leading to an integrated luminosity of $100\:\mathrm{fb}^{-1}$ per year and to a total integrated luminosity of 1 ab$^{-1}$ after 10 years of data-taking.

\section{High-energy cLFV observables at $\mu$TRISTAN}
\label{sec:muTobservables}

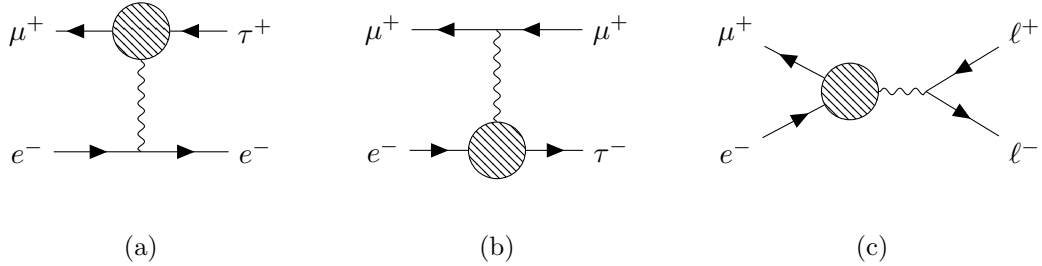
\begin{figure}[h!]
    \centering
    \begin{subfigure}[b]{0.24\textwidth}
    \centering
 \raisebox{1mm}{    \begin{tikzpicture}
    \begin{feynman}
    \vertex (a) at (0.5,0) {\(\mu^+\)};
    \vertex[blob] (b) at (2,0) {};
    \vertex (c) at (3.5,0) {\(\tau^+\)};
    \vertex (d) at (0.5,-1.6) {\(e^-\)};
    \vertex (e) at (2,-1.6) ;
    \vertex (f) at (3.5,-1.6)  {\(e^-\)};
    \diagram* {
    (a) -- [anti fermion] (b) -- [anti fermion] (c),
    (b) -- [boson] (e),
    (d) -- [fermion] (e) -- [fermion] (f)
    };
    \end{feynman}
    \end{tikzpicture}
    }
            \label{}
            \caption*{(a)}
    \end{subfigure} \hspace{3.5mm}
    \begin{subfigure}[b]{0.24\textwidth}
    \centering
 \raisebox{0.2mm}{    \begin{tikzpicture}
    \begin{feynman}
    \vertex (a) at (0.5,0) {\(\mu^+\)};
    \vertex (b) at (2,0);
    \vertex (c) at (3.5,0) {\(\mu^+\)};
    \vertex (d) at (0.5,-1.6) {\(e^-\)};
    \vertex [blob] (e) at (2,-1.6)  {};
    \vertex (f) at (3.5,-1.6)  {\(\tau^-\)};
    \diagram* {
    (a) -- [anti fermion] (b) -- [anti fermion] (c),
    (b) -- [boson] (e),
    (d) -- [fermion] (e) -- [fermion] (f)
    };
    \end{feynman}
    \end{tikzpicture}
    }
            \label{ }
            \caption*{(b)}
    \end{subfigure} \hspace{3.5mm}
    \begin{subfigure}[b]{0.24\textwidth}
    \centering
 \raisebox{1mm}{    \begin{tikzpicture}
    \begin{feynman}
    \vertex (a) at (0,0) {\(\mu^+\)};
    \vertex[blob] (b) at (1.5,-0.8) {};
    \vertex (c) at (3.8,0) {\(\ell^+\)};
    \vertex (d) at (0,-1.6) {\(e^-\)};
    \vertex (e) at (2.5,-0.8) ;
    \vertex (f) at (3.8,-1.6)  {\(\ell^-\)};
    \diagram* {
    (a) -- [anti fermion] (b) -- [anti fermion] (d),
    (b) -- [boson] (e),
    (c) -- [fermion] (e) -- [fermion] (f)
    };
    \end{feynman}
    \end{tikzpicture}
    }
            \label{}
            \caption*{\hspace{5mm}(c)}
    \end{subfigure}
    \caption{Feynman diagrams contributing to $\mu^+e^-\to \tau^+e^-$, $\mu^+e^-\to \mu^+\tau^-$ and $\mu^+e^-\to \ell^+\ell^-$. The shaded ``blob" denotes the inclusion of the loop diagrams depicted in Fig.~\ref{fig:triangle}.}
    \label{fig:diagram:penguin}
\end{figure}

\begin{figure}[h!]
    \centering
    \begin{subfigure}[b]{0.24\textwidth}
    \centering
 \raisebox{0mm}{    \begin{tikzpicture}
    \begin{feynman}
    \vertex (a) at (0.5,0) {\(\ell_\beta\)};
    \vertex (b) at (1.5,0);
    \vertex (c) at (3,0);
    \vertex (d) at (4,0) {\(\ell_\alpha\)};
    \vertex (e) at (2.25,-1) ;
    \vertex (f) at (2.25,-2) {\( \gamma/Z\)};
    \diagram* {
    (a) -- [fermion] (b),
    (b) -- [fermion, edge label=\(N_i\)] (c),
    (c) -- [fermion] (d),
    (b) -- [boson, edge label'=\(W\)] (e),
    (c) -- [boson, edge label=\(W\)] (e),
    (e) -- [boson]  (f)
    };
    \end{feynman}
    \end{tikzpicture}
    }
            \label{}
    \end{subfigure}
    \begin{subfigure}[b]{0.24\textwidth}
    \centering
 \raisebox{0.6mm}{    \begin{tikzpicture}
    \begin{feynman}
    \vertex (a) at (0.5,0) {\(\ell_\beta\)};
    \vertex (b) at (1.5,0);
    \vertex (c) at (3,0);
    \vertex (d) at (4,0) {\(\ell_\alpha\)};
    \vertex (e) at (2.25,-1) ;
    \vertex (f) at (2.25,-2)  {\(Z\)};
    \diagram* {
    (a) -- [fermion] (b),
    (b) -- [boson, edge label=\(W\)] (c),
    (c) -- [fermion] (d),
    (b) -- [fermion, edge label'=\(N_j\)] (e),
    (c) -- [anti fermion, edge label=\(N_i\)] (e),
    (e) -- [boson]  (f)
    };
    \end{feynman}
    \end{tikzpicture}
    }
            \label{}
    \end{subfigure}
    \begin{subfigure}[b]{0.24\textwidth}
    \centering
 \raisebox{4.7mm}{    \begin{tikzpicture}
    \begin{feynman}
    \vertex (a) at (0.5,0) {\(\ell_\beta\)};
    \vertex (b) at (1.5,0);
    \vertex (c) at (2,0);
    \vertex (d) at (2.75,0);
    \vertex (e) at (3.5,0) {\(\ell_\alpha\)};
    \vertex (f) at (1.5,-2) {\(\gamma/Z\)};
    \diagram* {
    (a) -- [fermion] (b),
    (b) -- [boson] (f),
    (b) -- [fermion] (c),
    (c) -- [fermion, half right,edge label'=\(N_i\)] (d),
    (c) -- [ half left,boson,edge label=\(W\)] (d),    
    (d) -- [fermion] (e)
    };
    \end{feynman}
    \end{tikzpicture}
    }
            \label{}
    \end{subfigure}
    \begin{subfigure}[b]{0.24\textwidth}
    \centering
 \raisebox{4.7mm}{    \begin{tikzpicture}
    \begin{feynman}
    \vertex (a) at (-0.5,0) {\(\ell_\alpha\)};
    \vertex (b) at (-1.5,0);
    \vertex (c) at (-2,0);
    \vertex (d) at (-2.75,0);
    \vertex (e) at (-3.5,0) {\(\ell_\beta\)};
    \vertex (f) at (-1.5,-2) {\(\gamma/Z\)};
    \diagram* {
    (a) -- [anti fermion] (b),
    (b) -- [boson] (f),
    (b) -- [anti fermion] (c),
    (c) -- [anti fermion, half left,edge label=\(N_i\)] (d),
    (c) -- [ half right,boson,edge label'=\(W\)] (d),    
    (d) -- [anti fermion] (e)
    };
    \end{feynman}
    \end{tikzpicture}
    }
            \label{}
    \end{subfigure}
    \caption{One-loop interactions contributing to the penguin-diagrams under consideration (see Fig.~\ref{fig:diagram:penguin}).}
    \label{fig:triangle}
\end{figure}
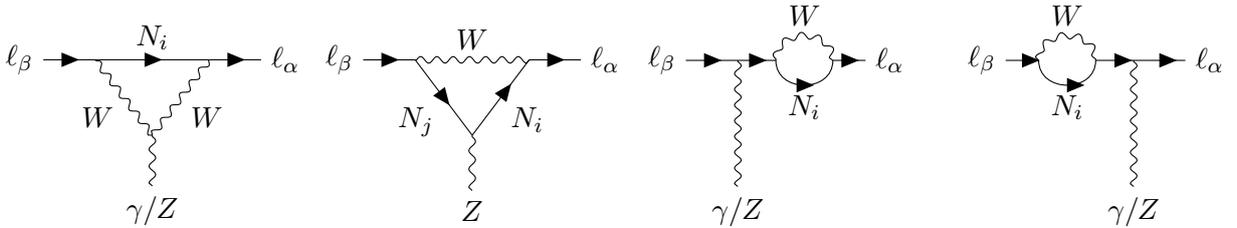

\begin{figure}[h!]
    \centering
    \begin{subfigure}[b]{0.4\textwidth}
    \centering
 \raisebox{-5mm}{    \begin{tikzpicture}
    \begin{feynman}
    \vertex (a) at (0,-0.5) {\(e^-\)};
    \vertex (b) at (2,0);
    \vertex (c) at (4,0);
    \vertex (d) at (6,-0.5) {\(\ell_\beta^-\)};
    \vertex (a1) at (0,2.5) {\(\mu^+\)};
    \vertex (b1) at (2,2);
    \vertex (c1) at (4,2);
    \vertex (d1) at (6,2.5) {\(\ell_\alpha^+\)};
    \diagram* {
    (a1) -- [anti fermion] (b1)-- [edge label'=\(N_i\)] (b)-- [anti fermion] (a),
    (d1) -- [fermion] (c1)-- [edge label=\(N_j\)] (c)-- [fermion] (d),
    (b1) -- [boson, edge label=\(W\)] (c1),
    (b) -- [boson, edge label'=\( W\)] (c),
    };
    \end{feynman}
    \end{tikzpicture}
    }
            \label{}
            \caption*{(1)}
    \end{subfigure} 
    \hspace*{12mm}
    \begin{subfigure}[b]{0.40\textwidth}
    \centering
 \raisebox{-5mm}{    \begin{tikzpicture}
    \begin{feynman}
    \vertex (a) at (0,-0.5) {\(e^-\)};
    \vertex (b) at (2,0);
    \vertex (c) at (4,0);
    \vertex (d) at (6,-0.5) {\(\ell_\beta^-\)};
    \vertex (a1) at (0,2.5) {\(\mu^+\)};
    \vertex (b1) at (2,2);
    \vertex (c1) at (4,2);
    \vertex (d1) at (6,2.5) {\(\ell_\alpha^+\)};
    \diagram* {
    (a1) -- [anti fermion] (b1)-- [edge label=\(N_i\)] (c1)-- [anti fermion] (d1),
    (a) -- [fermion] (b)-- [edge label'=\(N_j\)] (c)-- [fermion] (d),
    (b1) -- [boson, edge label'=\(W\)] (b),
    (c1) -- [boson, edge label=\( W\)] (c),
    };
    \end{feynman}
    \end{tikzpicture}
    }
            \label{}
            \caption*{(2)}
    \end{subfigure}
    \\
    \vspace*{5mm}
    \begin{subfigure}[b]{0.4\textwidth}
    \centering
 \raisebox{-5mm}{    \begin{tikzpicture}
    \begin{feynman}
    \vertex (a) at (0,-0.5) {\(e^-\)};
    \vertex (b) at (2,0);
    \vertex (c) at (4,0);
    \vertex (dt) at (4.8,1);
    \vertex (d) at (6,-0.5) {\(\ell_\beta^-\)};
    \vertex (a1) at (0,2.5) {\(\mu^+\)};
    \vertex (b1) at (2,2);
    \vertex (c1) at (4,2);
    \vertex (d1) at (6,2.5) {\(\ell_\alpha^+\)};
    \diagram* {
    (a1) -- [anti fermion] (b1)-- [edge label=\(N_i\)] (c1) -- []  (dt) -- [fermion]  (d),
    (a) -- [fermion] (b)-- [edge label'=\(N_j\)] (c) -- []  (dt)-- [anti fermion] (d1),
    (b1) -- [boson, edge label'=\(W\)] (b),
    (c1) -- [boson, edge label'=\( W\)] (c),
    };
    \end{feynman}
    \end{tikzpicture}
    }
            \label{}
            \caption*{(3)}
    \end{subfigure}
    \hspace*{12mm}
     \begin{subfigure}[b]{0.40\textwidth}
    \centering
 \raisebox{-5mm}{    \begin{tikzpicture}
    \begin{feynman}
    \vertex (a) at (0,-0.5) {\(e^-\)};
    \vertex (b) at (2,0);
    \vertex (c) at (4,0);
    \vertex (d) at (6,-0.5) {\(\ell_\beta^-\)};
    \vertex (a1) at (0,2.5) {\(\mu^+\)};
    \vertex (b1) at (2,1.7);
    \vertex (c1) at (4,1.7);
    \vertex (d1) at (6,2.5) {\(\ell_\alpha^+\)};
    \diagram* {
    (d1) -- [anti fermion] (b1)-- [edge label'=\(N_j\)] (b)-- [fermion] (a),
    (a1) -- [anti fermion] (c1)-- [edge label=\(N_i\)] (c)-- [fermion] (d),
    (b1) -- [boson, edge label'=\(W\)] (c1),
    (b) -- [boson, edge label'=\( W\)] (c),
    };
    \end{feynman}
    \end{tikzpicture}
    }
            \label{}
            \caption*{(4)}
    \end{subfigure}
    \caption{Box diagrams contributing to $\mu^+ e^-\to \ell^+_\alpha \ell^-_\beta$. The upper row only contains lepton number conserving interactions, while the lower row also contains lepton violating vertices.}
    \label{fig:boxes}
\end{figure}
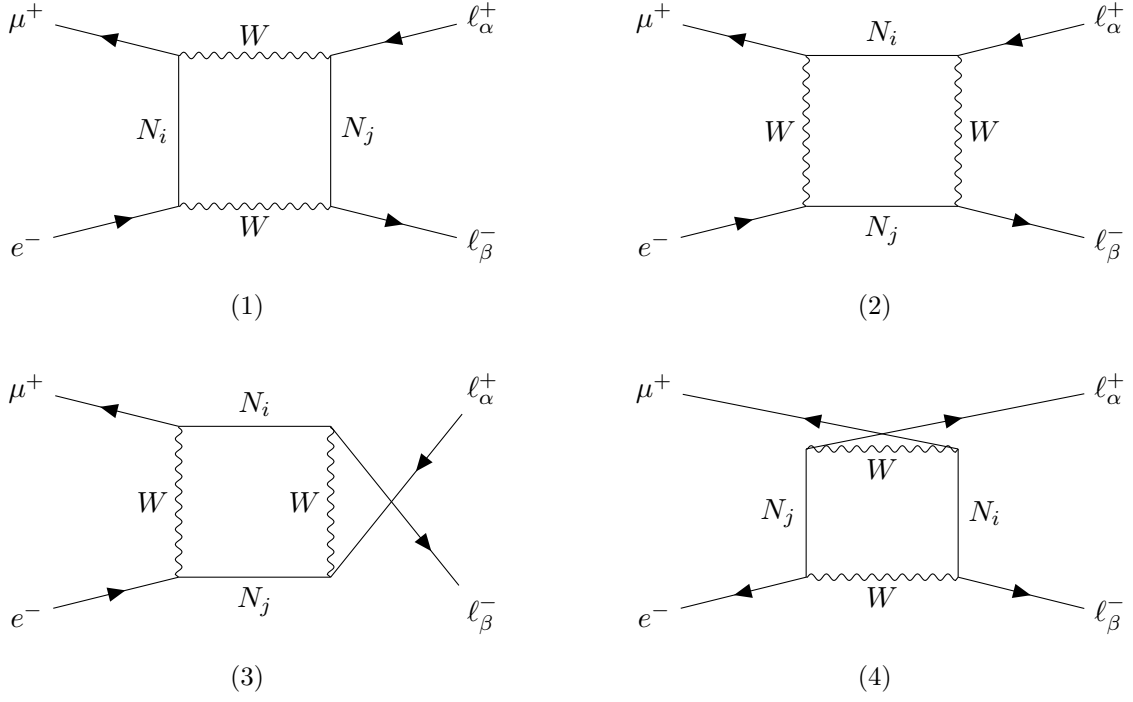

The flavour-asymmetric colliding beams allow to investigate a number of processes for which there is no SM contribution. These include, firstly
$\mu^+ e^- \to \tau^+ e^-$, relying on t-channel penguin processes (and flavour violating corrections to the outgoing lepton propagators) as well as box diagrams, in which the electron acts as the ``spectator flavour''. The contributing penguin diagrams  are depicted in Fig.~\ref{fig:diagram:penguin} (a). 
Secondly, one has $\mu^+ e^- \to \mu^+ \tau^-$, in full analogy to the previous case, but now with flavour violation occurring at the level of the electron line (spectator muon) (cf. Fig.~\ref{fig:diagram:penguin} (b)).
One also has $\mu^+ e^- \to \tau^+ \tau^-$ via s-channel and box diagrams, see Fig.~\ref{fig:diagram:penguin} (c) for the corresponding penguin contributions. Finally, one has the very interesting 
$\mu^+ e^- \to e^+ \mu^-$, reflecting a double flavour violation, and which only occur via the box diagrams depicted in Fig.~\ref{fig:boxes}. 
The relevant Feynman rules for these interactions as well as the interaction Lagrangian are summarised in Appendix~\ref{app:feynman}.

From the topology of the flavour violating processes previously discussed, it is manifest that there is a clear connection with high-energy cLFV $Z$ decays, and with several low-energy cLFV rare decays, including charged lepton radiative ($\ell_\alpha\to \ell_\beta \gamma$) and three-body ($\ell_\alpha\to \ell_\beta \ell_\delta\ell_\lambda$) decays, muonium oscillations as well as $\mu-e$ conversion - as we will further explore in subsequent sections.

\bigskip
As already mentioned, the amplitudes to the processes under consideration receive contributions from photon- and $Z$-penguins as well as box diagrams.
Before entering the computation of the amplitudes, we note that in the limit of massless external fermions and
once all Fierz transformations have been performed, the amplitudes admit only 3 types of Lorentz-structure: 
\begin{eqnarray}
    \mathcal M_{LL} &=&\mathcal A_{LL} \,[\bar v_\mu(p_1) \gamma^\mu P_L v_\alpha(p_3)]\,[\bar u_\beta(p_4) \gamma^\mu P_L u_e(p_2)]\,,\nonumber\\
    \mathcal M_{LR} &=& \mathcal A_{LR}\,[\bar v_\mu(p_1) \gamma^\mu P_L v_\alpha(p_3)]\,[\bar u_\beta(p_4) \gamma^\mu P_R u_e(p_2)]\,,\nonumber\\
    \mathcal M_{RL} &=& \mathcal A_{RL}\,[\bar v_\mu(p_1) \gamma^\mu P_R v_\alpha(p_3)]\,[\bar u_\beta(p_4) \gamma^\mu P_L u_e(p_2)]\,,
    \label{eqn:general_amps}
\end{eqnarray}
in which we chose the ordering of fermion fields as the reference order that all other fields will be Fierz-transformed into, with $\alpha\,,\beta = e\,,\mu\,,\tau$ depending on the given process\footnote{The final expressions to the amplitudes $\mathcal A_{LL,LR,RL}$ will be provided at the end of this section, Eqs.~(\ref{eq:firstA}-\ref{eq:lastA}). }. 
Squaring and spin-averaging the amplitude in the massless limit furthermore allows obtaining the differential cross-section as
\begin{equation}
    \frac{d\sigma}{d\cos\theta} = \frac{1}{32\pi \,s}\left(|\mathcal A_{LL}|^2 \,u^2 + (|\mathcal A_{LR}|^2 + |\mathcal A_{RL}|^2) \,s^2 \right)\,,\label{eqn:diffxsec}
\end{equation}
in which $\cos\theta = \frac{t - u}{s} = \frac{2t + s}{s}$ and $s\,,t\,,u$ are the usual Mandelstam variables.
Still in the massless limit, we have the following on-shell conditions
\begin{eqnarray}
    p_1^2 &=& p_2^2 \,= \, p_3^2 \,= \,p_4^2 = 0\,,\nonumber\\
    (p_1 + p_2)^2 &=& (p_3 + p_4)^2 \,= \,s\,,\nonumber\\
    (p_1 - p_3)^2 &=& (p_4 - p_2)^2 \,= \,t\,,\nonumber\\
    (p_1 - p_4)^2 &=& (p_4 - p_1)^2 \,= \,u\,,\nonumber\\
    s + t + u &=& 0\,,\nonumber\\
    (p_1 \,p_2) &=& (p_3 \,p_4)\, =\, \frac{s}{2}\,,\nonumber\\
    (p_1 \,p_3) &=& (p_2 \,p_4) \,= \,-\frac{t}{2}\,,\nonumber\\
    (p_1 \,p_4) &=& (p_2 \,p_3) \,=\, -\frac{u}{2}\,.
\end{eqnarray}
For each of the purely leptonic processes the contributions to the amplitudes $\mathcal A_{LL,LR,RL}$ exhibit (slight) differences.
We begin the discussion with the process $\mu^+ e^-\to \tau^+e^-$, noticing that the expressions for the other processes can be obtained with the appropriate re-arrangements of flavour indices and squared momenta (crossing symmetry).
We also stress that all diagrams have been computed in Feynman gauge~\footnote{Whilst the corresponding diagrams with Goldstone modes have been omitted from our figures, they are naturally included in our calculations.}. 

In the massless limit, the amplitudes of the photon- and $Z$-penguins can be cast as
\begin{eqnarray}
    \mathcal M_\gamma &=& -g \,s_w\, Q_e\,\frac{1}{t}\,F_\gamma^{\mu\tau}(t)\,[\bar v_\mu(p_1) \gamma^\mu P_L v_\tau(p_3)]\,[\bar u_\beta(p_4) \gamma^\mu u_e(p_2)]\,,\\
    \mathcal M_Z &=& -\frac{g}{c_w}\,\frac{1}{t-m_Z^2}\,F_Z^{\mu\tau}(t)\,[\bar v_\mu(p_1) \gamma^\mu P_L v_\tau(p_3)]\,[\bar u_\beta(p_4) \gamma^\mu(g_V^e -g_A^e \gamma_5) u_e(p_2)]\,,
\end{eqnarray}
in which $Q_e = -1$, the weak mixing angle encoded via $s_w, c_w = \sin\theta_w,\cos\theta_w$ (with $c_w = \frac{m_W}{m_Z}$ in the on-shell scheme), with the $Z$-couplings given as $g_V^f = \frac{1}{2} T_f^3 - Q_f \sin^2\theta_w$ and $g_A^f = \frac{1}{2}T_f^3$.
The form-factor $F_{\gamma}^{\alpha\beta}(q^2)$ can be  written as
\begin{eqnarray}
    F_\gamma^{\alpha\beta}(q^2) &=& -\frac{g^3\, s_w}{32 \pi^2 \,m_W^2}  \sum_{i} \mathcal U_{\alpha i}\,\mathcal U_{\beta i}^\ast \left[(m_i^2 + (D-4)m_W^2)B_0 + (m_i^2 + (D-2)m_W^2)\,B_1\right.\nonumber\\ 
    &\phantom{=}&\left.+ m_W^2\left(2(m_i^2 - m_W^2) \,C_0 + q^2\,(C_1 + C_2)\right) - 2(m_i^2 + (D-2)\,m_W^2)\,C_{00} \right]\,,
\end{eqnarray}
in which we abbreviated the Passarino-Veltman functions as $B_{0,1} \equiv B_{0,1}(0,m_i^2, m_W^2)$ and $C_{0,1,2,00} \equiv C_{0,1,2,00}(0,q^2,0,m_i^2, m_W^2, m_W^2)$, in the convention of LoopTools~\cite{Hahn:1998yk}; $D=4-2\epsilon$ is the dimensionality of space-time.
Furthermore, $\mathcal U$ denotes the extended lepton mixing matrix (see also Appendix~\ref{app:feynman}).
The squared vector-momentum $q^2$ has to be replaced with one of the Mandelstam variables $s,t,u$, depending on the process under consideration. For $\mu^+ e^-\to\tau^+ e^-$ we have $q^2 = t$.
Furthermore, the index $i$ runs over the internal neutrino states, while $\alpha,\beta$ are the charged lepton flavours (i.e. $\alpha = \mu$ and $\beta = \tau$ for $\mu^+e^-\to\tau^+e^-$).
The form-factor of the $Z$-penguin $F_{Z}^{\alpha\beta}(q^2)$ is  more involved and, for convenience, it is split into
\begin{eqnarray}
    F_Z^{\alpha\beta}(q^2) =  G_{Z}^{\alpha\beta}(q^2) + H_{Z}^{\alpha\beta}(q^2)\,,
\end{eqnarray}
with 
\begin{eqnarray}
    G_{Z}^{\alpha\beta}(q^2) &=& \frac{g^3}{32\pi^2 \,c_w \,m_W^2}\,\sum_i \mathcal U_{\alpha i}\,\mathcal U_{\beta i}^\ast\left[2 c_w^2 m_W^2 \,B_0 + (g_V^\ell + g_A^\ell)(m_i^2 + (D-2)\,m_W^2)\,(B_0 + B_1) \right. \nonumber\\
      &\phantom{=}&+\left. c_w m_W\left( 2(s_w^2\, m_i^2 \,m_Z + c_w \,m_W^3)\, C_0 - c_w \,m_W\, q^2 \,(C_1 + C_2)\right.\right.\nonumber\\
      &\phantom{=}&- \left.\left. ((2 c_w^2 - 1)\,m_i^2 + 2 \,c_w^2\,(D-2) \,m_W^2)\,C_{00}\right)\right]\,,
\end{eqnarray}
in which the Passarino-Veltman functions correspond to $B_{0,1} \equiv B_{0,1}(0,m_i^2, m_W^2)$ and $C_{0,1,2,00} \equiv C_{0,1,2,00}(0,q^2,0,m_i^2, m_W^2, m_W^2)$.
The second term is given by
\begin{eqnarray}
    H_Z^{\alpha\beta}(q^2) &=& \frac{g^3}{64\pi^2 \,c_w \,m_W^2}\,\sum_{i,j} \mathcal U_{\alpha i}\,\,\mathcal U_{\beta j}^\ast \left\{\mathcal C_{ij} \left[m_W^2\,((6-D) \,B_0^{ij} - 2 B_0^i - 2 B_0^j \right.\right.\nonumber\\
    &\phantom{=}&- \left.\left.(2m_i^2 + 2m_j^2 + (D-6) \,m_W^2 - 2 q^2)\,C_0 + 2(D-2)\,C_{00}) + m_i^2 \,m_j^2 \,C_0
    \right]\right.\nonumber\\
    &\phantom{=}&+\left.\mathcal C_{ij}^\ast m_i \,m_j\left[ B_0^{ij} - (D-3)\, m_W^2 \,C_0 -2 C_{00}\right]\right\}\,,
\end{eqnarray}
in which  $B_{0}^{i,j} \equiv B_{0}(0,m_W^2, m_i^2)$, $B_{0}^{ij} \equiv B_{0}(q^2, m_i^2, m_j^2)$,  and $C_{0,00} \equiv C_{0,00}(0,q^2,0,m_W^2, m_i^2, m_j^2)$ denote the compact version of the Passarino-Veltman functions.
For the definition of $\mathcal C$, see Eq.~\eqref{eq:Cij}.

Before proceeding to the box diagrams, let us stress here that while the $B_{0,1}$- and $C_{00}$-functions have divergent terms, their combinations entering in the full loop-functions are finite due to (semi-) unitarity of the lepton mixing matrix $\mathcal U$. 
For more details on the finiteness of the $Z$-penguin, see also the related discussion in~\cite{2207.10109}. 
Furthermore, we emphasise that the $Z$-penguin amplitude is only gauge-invariant for an on-shell $Z$, which is clearly not the case here.
Only upon inclusion of the corresponding photon-penguin as well as the box diagrams, is gauge-invariance  restored in the full amplitude. 
(This can be checked via an explicit computation of the process in the $R_\xi$-gauge, leading to extremely lengthy (analytical) expressions which we refrain from reproducing here.)
For a related discussion see e.g.~\cite{1903.05116}.

\medskip
While the computation of the penguin contributions is tedious but fairly straightforward, the computation of the box-diagrams contains several less trivial steps which are outlined in more detail in Appendix~\ref{app:boxes}.
Here, we just give the main results necessary to numerically evaluate the amplitude.
The four box diagrams shown in Figure~\ref{fig:boxes} contribute to $\mu^+ e^-\to \ell^+_\alpha \ell^-_\beta$ processes. We notice here that the upper row contains only Lepton Number conserving interactions while the lower row also contains LNV interactions or, more broadly speaking, fermion number violating vertices. The derivation of the Feynman rules for these vertices are taken from~\cite{Denner:1992vza} and summarised in Appendix~\ref{app:feynman}.

The four box contributions to the amplitude for $\mu^+ e^-\to \ell^+_\alpha \ell^-_\beta$ (following the labelling of Figure~\ref{fig:boxes}) can be written in the massless limit as
\begin{eqnarray}
    \mathcal M_1 &=& \mathcal B_1^{\alpha\beta}(t,s) \,[\bar v_\mu(p_1) \gamma^\mu P_L u_e(p_2)]\,[\bar u_\beta(p_4) \gamma^\mu P_L v_\alpha(p_3)]\,,\\
    \mathcal M_2 &=& \mathcal B_2^{\alpha\beta}(s,t) \,[\bar v_\mu(p_1) \gamma^\mu P_L v_\alpha(p_3)]\,[\bar u_\beta(p_4) \gamma^\mu P_L u_e(p_2)]\,,\\
    \mathcal M_3 &=& \mathcal B_3^{\alpha\beta}(s,u) \,[\bar v_\mu(p_1) P_R v_\beta(p_4)]\,[\bar u_\alpha(p_3)P_L u_e(p_2)]\,,\\
    \mathcal M_4 &=& \mathcal B_4^{\alpha\beta}(t,u) \,[\bar v_\mu(p_1) P_R v_\beta(p_4)]\,[\bar u_\alpha(p_3)P_L u_e(p_2)]\,.
\end{eqnarray}
Clearly, the four-fermion structures appearing in $\mathcal M_1, \mathcal M_3$ and $\mathcal M_4$ have to be Fierz-transformed to the reference order of the penguins, which coincides with the ordering of $\mathcal M_2$. Making use of the Fierz identities derived in~\cite{Nieves:2003in}, one can show that 
\begin{eqnarray}
    [\bar v_\mu(p_1) \gamma^\mu P_L u_e(p_2)]\,[\bar u_\beta(p_4) \gamma^\mu P_L v_\alpha(p_3)] &=& [\bar v_\mu(p_1) \gamma^\mu P_L v_\alpha(p_3)]\,[\bar u_\beta(p_4) \gamma^\mu P_L u_e(p_2)]\,,\label{eqn:fierz1}\\
    {[\bar v_\mu(p_1) P_R v_\beta(p_4)]\,[\bar u_\alpha(p_3)P_L u_e(p_2)]} &=& -\frac{1}{2}[\bar v_\mu(p_1) \gamma^\mu P_L v_\alpha(p_3)]\,[\bar u_\beta(p_4) \gamma^\mu P_L u_e(p_2)]\,.\label{eqn:fierz2}
\end{eqnarray}
The various box form-factors and their loop-functions can be written as
\begin{eqnarray}
    \mathcal B_1^{\alpha\beta}
(t,s) &=& \frac{g^4}{128\pi^2 }\sum_{i,j} \mathcal U_{\alpha j}^\ast\,\mathcal U_{\beta j}\,\mathcal U_{e i}^\ast\,\mathcal U_{\mu i}\,F_1(t,s)\,,\\
\mathcal B_2^{\alpha\beta}(s,t) &=& \frac{g^4}{128\pi^2 }\sum_{i,j} \mathcal U_{\alpha i}^\ast\,\mathcal U_{\beta j}\,\mathcal U_{e j}^\ast\,\mathcal U_{\mu i}\,F_1(s,t)\,,\\
\mathcal B_3^{\alpha\beta}(s,u) &=& \frac{g^4}{64\pi^2}\sum_{i,j} \mathcal U_{\alpha j}^\ast\,\mathcal U_{\beta i}\,\mathcal U_{e j}^\ast\,\mathcal U_{\mu i}\,F_2(s,u)\,,\\
\mathcal B_4^{\alpha\beta}(t,u) &=& \frac{g^4}{64\pi^2}\sum_{i,j} \mathcal U_{\alpha j}^\ast\,\mathcal U_{\beta i}\,\mathcal U_{e j}^\ast\,\mathcal U_{\mu i}\,F_2(t,u)\,,\\
F_1 (q_1^2, q_2^2) &=&\frac{1}{m_W^4}\bigg(q_1^2\, (m_i^2 m_j^2 + 4 m_W^4)\,(D_{1} + D_{11} + D_{12} + D_{13})\nonumber\\
&\phantom{=}& + 4 m_W^4 \,q_1^2\,(D_2 + D_3) - 4 m_i^2 \,m_j^2\, m_W^2 \,D_0 + 2(m_i^2 \,m_j^2 + 4 m_W^4)\,D_{00}\bigg)\,,\\
F_2(q_1^2, q_2^2) &=& \frac{m_i m_j}{m_W^4}\bigg( (m_i^2\, m_j^2 + 4 m_W^4)\,D_0 - 2 m_W^2\Big(4 D_{00}  \nonumber\\
&\phantom{=}&+ 2 q_1^2\,(D_{1} + D_{11} + D_{12} + D_{13} + D_{23} + \frac{D_{2} + D_{3}}{2})\Big)\bigg)\,,
\end{eqnarray}
in which the Passarino-Veltman four-point functions (in the LoopTools~\cite{Hahn:1998yk} convention) were introduced as
$D_{ij} \equiv D_{ij}(q_1^2, 0, q_2^2, 0, 0, 0, m_i^2, m_j^2, m_W^2, m_W^2)$. We note that the four-point functions are all finite.

\medskip
We are now able to bring together all contributions to the amplitudes defined in~\eqref{eqn:general_amps}.
For the process $\mu^+ e^-\to \tau^+ e^-$ they are obtained in a straightforward manner from the penguins and boxes: making use of the Fierz rearrangements in Eq.~\eqref{eqn:fierz1} and~\eqref{eqn:fierz2}, as
\begin{eqnarray}
    \mathcal A_{LL}^{\mu e\tau e} &=& \mathcal B_1^{\tau e}(t,s) + \mathcal B_2^{\tau e}(s,t) - \frac{\mathcal B_3^{\tau e}(s,u) + \mathcal B_4^{\tau e}(t,u)}{2} - \frac{g}{c_w}\frac{g_V^e + g_A^e}{t - m_Z^2}\, F_Z^{\mu\tau}(t) - \frac{g \,s_w\, Q_e}{t} \,F_\gamma^{\mu\tau}(t)\,,\\
    \mathcal A_{LR}^{\mu e\tau e} &=& - \frac{g}{c_w}\frac{g_V^e - g_A^e}{t - m_Z^2}\,F_Z^{\mu\tau}(t) - \frac{g \,s_w \,Q_e}{t} \,F_\gamma^{\mu\tau}(t)\,,\\
    \mathcal A_{RL}^{\mu e\tau e} &=& 0\,.
\end{eqnarray}

\medskip
For the other processes, the corresponding expressions can be derived simply by appropriately re-arranging flavour indices and momentum transfers.
For $\mu^+ e^-\to \mu^+ \tau^-$ we have the same box-contributions but with the exchange $\tau\leftrightarrow e$ indices, while the penguin contributions require  more work. Instead of having the flavour violating penguin insertion in the muon line (see diagram (a) in Figure~\ref{fig:diagram:penguin}), it now appears in the electron line (see  Figure~\ref{fig:diagram:penguin}~(b)); after some algebra this leads to
\begin{eqnarray}
    \mathcal A_{LL}^{\mu e\mu \tau} &=& \mathcal B_1^{e\tau}(t,s) + \mathcal B_2^{e\tau}(s,t) - \frac{\mathcal B_3^{e\tau}(s,u) + \mathcal B_4^{e\tau}(t,u)}{2} - \frac{g}{c_w}\frac{g_V^\mu + g_A^\mu}{t - m_Z^2} F_Z^{\tau e}(t) - \frac{g \,s_w\, Q_\mu}{t}\, F_\gamma^{\tau e}(t)\,, \label{eq:firstA}\\
    \mathcal A_{LR}^{\mu e\mu \tau} &=& 0\,,\\
    \mathcal A_{RL}^{\mu e\mu \tau} &=& - \frac{g}{c_w}\frac{g_V^\mu - g_A^\mu}{t - m_Z^2}\,F_Z^{\tau e}(t) - \frac{g \,s_w \,Q_\mu}{t}\, F_\gamma^{\tau e}(t)\,.
\end{eqnarray}

\medskip
For the third case of $\mu^+ e^-\to \ell^+\ell^-$ pair-production, we need to distinguish 3 sub-cases depending on final state flavours.
In addition to the t-channel like penguin contributions we also have now annihilation s-channel like penguins (see diagram (c) of Figure~\ref{fig:diagram:penguin}).
These lead to different four-fermion structures which can however  be easily Fierz-rearranged into the reference order via Eq.~\eqref{eqn:fierz1} and Eq.~\eqref{eqn:fierz2}.
The amplitudes consequently receive contributions from the boxes, s-channel penguins and in the cases of $\ell = e(\mu)$ also ``upper''(``lower'') t-channel penguins.
These are given by
\begin{eqnarray}
    \mathcal A_{LL}^{\mu e ee} &=& \mathcal B_1^{ee}(t,s) + \mathcal B_2^{ee}(s,t) - \frac{\mathcal B_3^{ee}(s,u) + \mathcal B_4^{ee}(t,u)}{2} - \frac{g}{c_w}\frac{g_V^e + g_A^e}{t - m_Z^2}\, F_Z^{\mu e}(t) - \frac{g \,s_w\, Q_e}{t} \,F_\gamma^{\mu e}(t)\nonumber\label{eqn:amppair1}\\
    &\phantom{=}& - \frac{g}{c_w}\frac{g_V^e + g_A^e}{s - m_Z^2}\,F_Z^{\mu e}(s) - \frac{g \,s_w\, Q_e}{s} \,F_\gamma^{\mu e}(s)\,,\\
    \mathcal A_{LR}^{\mu e e e} &=& - \frac{g}{c_w}\frac{g_V^e - g_A^e}{t - m_Z^2}\,F_Z^{\mu e}(t) - \frac{g \,s_w\, Q_e}{t}\, F_\gamma^{\mu e}(t) - \frac{g}{c_w}\frac{g_V^e - g_A^e}{s - m_Z^2}\,F_Z^{\mu e}(s) - \frac{g \,s_w \,Q_e}{s} \,F_\gamma^{\mu e}(s)\,,\\
    \mathcal A_{RL}^{\mu e e e} &=& 0\,,\\
     \mathcal A_{LL}^{\mu e \mu\mu} &=& \mathcal B_1^{\mu\mu}(t,s) + \mathcal B_2^{\mu\mu}(s,t) - \frac{\mathcal B_3^{\mu\mu}(s,u) + \mathcal B_4^{\mu\mu}(t,u)}{2} - \frac{g}{c_w}\frac{g_V^\mu + g_A^\mu}{t - m_Z^2} \,F_Z^{\mu e}(t) - \frac{g \,s_w \,Q_\mu}{t} \,F_\gamma^{\mu e}(t)\nonumber\\
    &\phantom{=}& - \frac{g}{c_w}\frac{g_V^\mu + g_A^\mu}{s - m_Z^2} \,F_Z^{\mu e}(s) - \frac{g \,s_w \,Q_\mu}{s} \,F_\gamma^{\mu e}(s)\,,\\
    \mathcal A_{LR}^{\mu e \mu \mu} &=&  - \frac{g}{c_w}\frac{g_V^\mu - g_A^\mu}{s - m_Z^2}\,F_Z^{\mu e}(s) - \frac{g \,s_w \,Q_\mu}{s}\, F_\gamma^{\mu e}(s)\,,\\
    \mathcal A_{RL}^{\mu e \mu \mu} &=& - \frac{g}{c_w}\frac{g_V^\mu - g_A^\mu}{t - m_Z^2}\,F_Z^{\mu e}(t) - \frac{g\,s_w\, Q_\mu}{t}\, F_\gamma^{\mu e}(t)\,,\label{eqn:amppair2}\\
    \mathcal A_{LL}^{\mu e \tau\tau} &=& \mathcal B_1^{\tau\tau}(t,s) + \mathcal B_2^{\tau\tau}(s,t) - \frac{\mathcal B_3^{\tau\tau}(s,u) + \mathcal B_4^{\tau\tau}(t,u)}{2} - \frac{g}{c_w}\frac{g_V^\tau + g_A^\tau}{s - m_Z^2} \,F_Z^{\mu e}(s) - \frac{g \,s_w\, Q_\tau}{s} \,F_\gamma^{\mu e}(s)\\
     \mathcal A_{LR}^{\mu e \tau \tau} &=&  - \frac{g}{c_w}\frac{g_V^\tau - g_A^\tau}{s - m_Z^2}\,F_Z^{\mu e}(s) - \frac{g \,s_w \,Q_\tau}{s}\, F_\gamma^{\mu e}(s)\,,\\
     \mathcal A_{RL}^{\mu e \tau \tau} &=& 0\,.
\end{eqnarray}

\medskip
Lastly, we have the high-energy equivalent of muonium-antimuonium oscillation, namely $\mu^+ e^- \to e^+\mu^-$. 
Just like its low-energy counterpart, it is only mediated by box diagrams and thus we have 
\begin{eqnarray}
     \mathcal A_{LL}^{\mu e e\mu} &=& \mathcal B_1^{e\mu}(t,s) + \mathcal B_2^{e\mu}(s,t) - \frac{\mathcal B_3^{e\mu}(s,u) + \mathcal B_4^{e\mu}(t,u)}{2}\,,\\
     \mathcal A_{LR}^{\mu ee\mu} &=& \mathcal A_{RL}^{\mu ee \mu} = 0\,.\label{eq:lastA}
\end{eqnarray}

\section{Theoretical frameworks and experimental constraints}\label{sec:HNL:model:constraints}

As discussed in the Introduction, heavy sterile states are naturally present in numerous well-motivated models of New Physics, especially those aiming at addressing neutrino mass generation. Many such Beyond the Standard Model (BSM) realisations offer rich phenomenological signatures, which in turn lead to extensive constraints on these new states. Below we briefly describe the theoretical frameworks upon which we will rely for our studies, also summarising the most pertinent constraints (phenomenological and experimental) on heavy neutral leptons.

\subsection{SM extensions via HNL}
As mentioned before, in order to illustrate the most relevant aspects of our study, we choose to consider two SM extensions featuring HNL: a minimal ``3+2'' construction and a realisation of the Inverse Seesaw, the ISS(3,3).

\paragraph{Ad-hoc minimal construction}
A possible approach to study the role of HNL in cLFV processes consists in considering minimal extensions of the SM, in which $n_S$ sterile states are added to the particle content. Other than assuming that these heavy neutral fermions are massive (and of Majorana nature), and that they do have non-negligible mixings to the active neutrinos, no other hypothesis is made, in particular concerning their contribution to the mechanism of neutrino mass generation. 
In these minimal ``ad-hoc'' models the (generalised) leptonic mixings are thus encoded in a $(3+n_S)\times(3+n_S)$ unitary mixing matrix, $\mathcal{U}$; its upper left $3\times3$ block corresponds to the left-handed leptonic mixing matrix (the would-be PMNS,  $\tilde{U}_\text{PMNS}$) - no longer unitary.

In what follows we will consider the case of $n_S=2$: the neutral lepton spectrum is thus composed of three light and 2 heavy mass eigenstates ($m_i$, with $i=1...5$); the $5 \times 5$ unitary leptonic mixing matrix can be parametrised via ten angles (through subsequent rotations, also comprising 6 Dirac phases) and a diagonal matrix including four physical Majorana phases (cf.~\cite{Abada:2015trh,2107.06313}).

The enlarged lepton mixing matrix $\mathcal U$ can be parametrised in the Euler representation via ten subsequent rotations $R_{ij}$ (with $i\neq j$), multiplied with a diagonal matrix including the four physical Majorana phases $\varphi_i$
\begin{eqnarray}
    \mathcal{U} \,= \,R_{45}\,R_{35}\,R_{25}\,R_{15}\,
    R_{34}\,R_{24}\,R_{14}\,R_{23}\,R_{13}\,R_{12}\times\mathrm{diag}(1, e^{i\varphi_2}, e^{i\varphi_3}, e^{i\varphi_4}, e^{i\varphi_5})\,.
    \label{eqn:allrot}
\end{eqnarray}
The rotations around the Euler angles are given by (illustrated by $R_{24}$):
\begin{equation}\label{eq:R24}
     R_{24} = \begin{pmatrix}
                 1 & 0 & 0 & 0 & 0\\
                 0 & \cos\theta_{24} & 0 & \sin \theta_{24} e^{-i\delta_{24}} & 0\\
                 0 & 0 & 1 & 0 & 0\\
                 0 & -\sin \theta_{24} e^{-i\delta_{24}} & 0 & \cos\theta_{24} & 0\\
                 0 & 0 & 0 & 0 & 1
             \end{pmatrix}\,.
 \end{equation}
We further note that the angles of the first three rotations ($R_{12}$, $R_{13}$ and $R_{23}$) are fixed by neutrino oscillation data~\cite{Esteban:2024eli}.
Finally, one has to choose 6 independent linear combinations of the ten ``Dirac-like'' CP phases present in the rotation matrices as physical.

\paragraph{A simple Inverse Seesaw realisation}
The Inverse Seesaw constitutes a variant of the type I seesaw which allows for a natural explanation of the smallness of the observed neutrino masses. It relies on the introduction of two species of sterile fermions, $X$ and $\nu_R$. The new terms at the origin of neutrino mass generation can be cast as
\begin{equation}\label{eq:ISS:lagrangian}
    \mathcal L_\text{ISS}\, =\, -Y^D_{ij} \,\overline{L_i^c}\,\widetilde H \,\nu_{Rj}^c - M_R^{ij}\, \overline{\nu_{Ri}}\, X_j - \frac{1}{2}\mu_R^{ij}\, \overline{\nu_{Ri}^c}\,\nu_{Rj} - \frac{1}{2} \mu_X^{ij}\, \overline{X_i^c}\, X_j + \text{H.c.}\,.
\end{equation}
Notice that upon $\mu_{X,R} \to 0$, one recovers total lepton number conservation. 
In the limit of approximate lepton number conservation, 
and taking for simplicity $\mu_{R} \to 0$ i.e. $\mu_X \ll m_D\ll M_R$, the masses of the light neutrinos are, to a good approximation\footnote{Notice that $\mu_R$ only appears at higher orders in the seesaw expansion in the active neutrino mass and can therefore be safely neglected, see for example~\cite{Dev:2012sg}.}, given by 
\begin{equation}
    m_\nu \simeq  m_D \, \left( M_{R}^{-1} \right)^{T} \, \mu_X \, M_{R}^{-1} \, m_D^T\, \equiv U_\text{PMNS}^\ast\, m_\nu^\text{diag}\, U_\text{PMNS}^\dagger\,.
    \label{eq:ISS:lightmasses}
\end{equation}
As can be seen from the above expression, the limit $\mu_{X} \to 0$ (and hence the generation of small neutrino masses for not excessively heavy $M_R$) becomes natural in the sense of  
't Hooft~\cite{tHooft:1980xss,Hettmansperger:2011bt}, since 
for $\mu_X=0$ lepton number is restored. 

In what follows we will consider a simple realisation of the ISS in which $n_R = n_X = 3$ generations of heavy sterile fermions are added to SM particle content, the so-called ISS(3,3)\footnote{For a detailed discussion of the most minimal ISS realisations, see~\cite{Abada:2014vea}. These include the ISS(2,2) and the ISS(2,3). It is also important to highlight that other well-motivated low-energy seesaw realisations include the Linear Seesaw~\cite{Akhmedov:1995vm,Barr:2003nn,Malinsky:2005bi}; its most minimal realisation relies on four additional states.}. 

As frequently done upon phenomenological studies of BSM scenarios of neutrino mass generation, it proves convenient and useful to consider parametrisations of the flavour structures of the model. 
For the ISS, 
a possibility consists in a modified Casas-Ibarra parametrisation~\cite{Casas:2001sr}, 
\begin{equation}
    m_D^T = V^\dagger\sqrt{M^\text{diag}}R\sqrt{m_\nu^\text{diag}} U_\text{PMNS}^\dagger\,,
    \label{eqn:ISSCI}
\end{equation}
in which the Dirac mass term (or equivalently the Yukawa couplings) encodes neutrino data, with the additional degrees of freedom parametrised via a complex orthogonal matrix $R$; in the above equation, the unitary matrix 
$V$ diagonalises $M = M_R \mu_X^{-1} M_R^T$ as $M = V^\dagger M^\text{diag}V^\ast$, and $R$ can be parametrised as
\begin{equation}
    R = \begin{pmatrix} c_2 c_3 & -c_1 s_3 - s_1 s_2 c_3 & s_1 s_3 - c_1 s_2 c_3\\
                        c_2 s_3 & c_1 c_3 - s_1 s_2 s_3 & -s_1 c_3 - c_1 s_2 s_3\\
                        s_2 & s_1 c_2 & c_1 c_2
        \end{pmatrix}\,,
        \label{eqn:Rmatrix}
\end{equation}
in which $s_i \equiv \sin\theta_i$, $c_i \equiv \cos\theta_i$ and $\theta_i\in \mathbb{C}$.

\subsection{Constraints on heavy neutral leptons}
In association with the potentially light new states, the above described modified currents open the door to contributions to numerous observables, ranging from flavoured transitions (including cLFV observables), to  
electroweak precision observables, among others. Likewise, 
neutrino oscillation data are also taken into account (using the recent results of NuFit6.0~\cite{Esteban:2024eli}).

\paragraph{Phenomenological constraints: perturbative unitarity}

Bounds on the width of the heavy states can be inferred from constraints arising from perturbative unitarity\footnote{For a recent explicit computation of all relevant partial wave amplitudes in the $2\to2$-scatterings see~\cite{Urquia-Calderon:2024rzc}.}~\cite{Chanowitz:1978mv,Durand:1989zs,Bernabeu:1993up,Fajfer:1998px,Ilakovac:1999md}:
\begin{equation}
    \frac{\Gamma(N_i)}{m_{N_i}}<\frac{1}{2},\quad \text{for}\quad i\geq4\,;
\end{equation}
for an HNL spectrum considerably heavier than all SM states (as is our case), the leading contributions to the HNL decay width arise from the two-body decays into a SM boson and a lepton. At tree level, these are given by (see, e.g.~\cite{Abada:2023raf})
\begin{eqnarray}
    \Gamma\left(N_i\rightarrow Wl_{\alpha}\right)&=&\frac{g_w^2}{64\pi}\left|\mathcal{U}_{\alpha i}\right|^2\frac{\lambda^{1/2}(m_i^2,M_W^2,m_{\ell_{\alpha}}^2)}{m_i}\left\lbrace 1+\frac{m_{\ell_{\alpha}}^2-2M_W^2}{m_i^2}+\frac{(m_i^2-m_{\ell_{\alpha}}^2)^2}{m_i^2M_W^2}\right\rbrace,\\
    \Gamma\left(N_i\rightarrow ZN_j\right)&=&\frac{g_w^2}{128c_w^2\pi}\left|C_{ij}\right|^2\frac{\lambda^{1/2}(m_i^2,M_Z^2,m_{j}^2)}{m_i}\left\lbrace 1+\frac{m_{j}^2-2M_Z^2}{m_i^2}+\frac{(m_i^2-m_{j}^2)^2}{m_i^2M_Z^2}\right\rbrace,\\
    \Gamma\left(N_i\rightarrow HN_j\right)&=&\frac{g_w^2}{128\pi}\left|C_{ij}\right|^2\frac{m_i^2}{M_W^2}\frac{\lambda^{1/2}(m_i^2,M_H^2,m_j^2)}{m_i}\left\lbrace\left(1+\frac{m_j^2-M_H^2}{m_i^2}\right)\left(1+\frac{m_j^2}{m_i^2}\right)+4\frac{m_j^2}{m_i^2}\right\rbrace, \nonumber
    \\
\end{eqnarray}
in which $\lambda(a,b,c)=\left[a-(\sqrt{b}-\sqrt{c})^2\right]\left[a-(\sqrt{b}+\sqrt{c})^2\right]$ is the K\"all\'en function. In turn, the constraints on the HNL decay width will have an impact on the allowed parameter space of the considered theoretical framework. 

\paragraph{A non-unitary leptonic mixing matrix}
The deviations of the $U_\text{PMNS}$ from unitarity (induced by the mixings between active and sterile neutral fermions) are at the origin of deviations from the SM expectations - concerning both flavour violating and flavour conserving leptonic transitions. 
The so-called $\eta$-matrix, defined as~\cite{FernandezMartinez:2007ms}
\begin{equation}
\label{eq:defPMNSeta}
U_\text{PMNS} \, \to \, \tilde U_\text{PMNS} \, = \,(\mathbb{1} - \eta)\, 
U_\text{PMNS}\,,
\end{equation}
proves convenient to evaluate deviations of $U_\text{PMNS}$ from unitarity. Comprehensive analyses have allowed to infer constraints on the entries of $\eta$~\cite{Fernandez-Martinez:2015hxa,Fernandez-Martinez:2016lgt,Blennow:2023mqx,Abada:2023raf}, stemming from a number of flavour observables and electroweak precision tests: these include oblique parameters, the invisible width of the $Z$ boson, muon decays, 
among others. 
Since these observables lead to subdominant constraints in the numerical analysis, we will not discuss them in detail, rather focusing on the new contributions to flavoured observables.

\paragraph{Flavoured observables: charged lepton flavour violation and lepton flavour universality}
In general, constraints from the non-observation of rare cLFV transitions and decays are at origin of the strongest bounds on SM extensions via HNL states. Among the most important modes one has cLFV three-body  decays, $\mu - e$ conversion in muonic atoms, as well as cLFV $Z$ boson decays. The bounds and future sensitivities concerning the decays which will be considered in our study are summarised in Table~\ref{tab:cLFV_lep}.
In the context of SM extensions via $n_S$ generations of HNL, the expressions for the cLFV observables can be found, for example, in~\cite{Ilakovac:1994kj,Alonso:2012ji,Abada:2018nio,Abada:2022asx,Riemann:1982rq,Illana:1999ww,Mann:1983dv,Illana:2000ic,Ma:1979px,Gronau:1984ct,Deppisch:2004fa,Deppisch:2005zm,Dinh:2012bp,Abada:2014kba,Abada:2015oba,Abada:2015zea,Abada:2016vzu,Arganda:2014dta}.
\renewcommand{\arraystretch}{1.3}
\begin{table}[h!]
    \centering
    \hspace*{-2mm}{\small\begin{tabular}{|c|c|c|}
    \hline
    Observable & Current bound & Future sensitivity  \\
    \hline\hline
    $\text{BR}(\mu\to e \gamma)$    &
    \quad $<3.1\times 10^{-13}$ \quad (MEG II~\cite{MEGII:2023ltw})   &
    \quad $6\times 10^{-14}$ \quad (MEG II~\cite{Baldini:2018nnn}) \\
    $\text{BR}(\tau \to e \gamma)$  &
    \quad $<3.3\times 10^{-8}$ \quad (BaBar~\cite{Aubert:2009ag})    &
    \quad $3\times10^{-9}$ \quad (Belle II~\cite{Kou:2018nap})      \\
    $\text{BR}(\tau \to \mu \gamma)$    &
     \quad $ <4.2\times 10^{-8}$ \quad (Belle~\cite{Belle:2021ysv})  &
    \quad $10^{-9}$ \quad (Belle II~\cite{Kou:2018nap})     \\
    \hline
    $\text{BR}(\mu \to 3 e)$    &
     \quad $<1.0\times 10^{-12}$ \quad (SINDRUM~\cite{Bellgardt:1987du})    &
     \quad $10^{-15(-16)}$ \quad (Mu3e~\cite{Blondel:2013ia})   \\
    $\text{BR}(\tau \to 3 e)$   &
    \quad $<2.7\times 10^{-8}$ \quad (Belle~\cite{Hayasaka:2010np})&
    \quad $5\times10^{-10}$ \quad (Belle II~\cite{Kou:2018nap})     \\
    $\text{BR}(\tau \to 3 \mu )$    &
    \quad $<1.9\times 10^{-8}$ \quad (Belle II~\cite{Belle-II:2024sce})  &
    \quad $5\times10^{-10}$ \quad (Belle II~\cite{Kou:2018nap})     \\
    & & \quad$5\times 10^{-11}$\quad (FCC-ee~\cite{Abada:2019lih})\\
        $\text{BR}(\tau^- \to e^-\mu^+\mu^-)$   &
    \quad $<2.7\times 10^{-8}$ \quad (Belle~\cite{Hayasaka:2010np})&
    \quad $5\times10^{-10}$ \quad (Belle II~\cite{Kou:2018nap})     \\
    $\text{BR}(\tau^- \to \mu^-e^+e^-)$ &
    \quad $<1.8\times 10^{-8}$ \quad (Belle~\cite{Hayasaka:2010np})&
    \quad $5\times10^{-10}$ \quad (Belle II~\cite{Kou:2018nap})     \\
    $\text{BR}(\tau^- \to e^-\mu^+e^-)$ &
    \quad $<1.5\times 10^{-8}$ \quad (Belle~\cite{Hayasaka:2010np})&
    \quad $3\times10^{-10}$ \quad (Belle II~\cite{Kou:2018nap})     \\
    $\text{BR}(\tau^- \to \mu^-e^+\mu^-)$   &
    \quad $<1.7\times 10^{-8}$ \quad (Belle~\cite{Hayasaka:2010np})&
    \quad $4\times10^{-10}$ \quad (Belle II~\cite{Kou:2018nap})     \\
    \hline
    $\text{CR}(\mu- e, \text{N})$ &
     \quad $<7 \times 10^{-13}$ \quad  (Au, SINDRUM~\cite{Bertl:2006up}) &
    \quad $10^{-14}$  \quad (SiC, DeeMe~\cite{Nguyen:2015vkk})    \\
    & &  \quad $2.6\times 10^{-17}$  \quad (Al, COMET~\cite{Krikler:2015msn,COMET:2018auw,Moritsu:2022lem})  \\
    & &  \quad $8 \times 10^{-17}$  \quad (Al, Mu2e~\cite{Bartoszek:2014mya})\\
    \hline
    P(Mu-$\overline{\mathrm{Mu}}$) & $< 8.3\times 10^{-11}$ \quad (PSI~\cite{Willmann:1998gd}) \quad&  \quad$\mathcal O (10^{-13})$\quad (MACE~\cite{Bai:2024skk})
    \\ 
    \hline
    $\mathrm{BR}(Z\to e^\pm\mu^\mp)$ & \quad$< 4.2\times 10^{-7}$\quad (ATLAS~\cite{Aad:2014bca}) & \quad$\mathcal O (10^{-10})$\quad (FCC-ee~\cite{Abada:2019lih})\\
    $\mathrm{BR}(Z\to e^\pm\tau^\mp)$ & \quad$< 4.1\times 10^{-6}$\quad (ATLAS~\cite{ATLAS:2021bdj}) & \quad$\mathcal O (10^{-10})$\quad (FCC-ee~\cite{Abada:2019lih})\\
    $\mathrm{BR}(Z\to \mu^\pm\tau^\mp)$ & \quad$< 5.3\times 10^{-6}$\quad (ATLAS~\cite{ATLAS:2021bdj}) & \quad $\mathcal O (10^{-10})$\quad (FCC-ee~\cite{Abada:2019lih})\\
    \hline
    \end{tabular}}
    \caption{Current experimental bounds and future sensitivities on relevant cLFV observables where BR denotes the branching ratio, CR the conversion rate and P is the oscillation probability. The quoted limits are given at $90\%\:\mathrm{C.L.}$ (Belle II sensitivities correspond to an integrated luminosity of $50\:\mathrm{ab}^{-1}$.)}
    \label{tab:cLFV_lep}
\end{table}
\renewcommand{\arraystretch}{1.}

The modified charged and neutral currents (a direct consequence of the non-unitarity of the PMNS mixing matrix) also has non-negligible consequences for (flavoured) universality ratios. This is the case for leptonic decays of light mesons - such as kaons and pions (see, for example~\cite{Abada:2013aba,Abada:2012mc}), as well as tau-lepton decays from the comparison of the 
widths $\Gamma(\tau \rightarrow \mu \nu \nu)$ and $\Gamma(\tau \rightarrow e \nu \nu)$ (i.e. the $R_{\tau}$ ratio).
 Likewise, and at the frontier with EWPO, one also must also take into account lepton flavour universality tests in $Z$ boson decays, 
 $\Gamma(Z \rightarrow \ell_\alpha \ell_\alpha)$ and $\Gamma(Z \rightarrow \ell_\beta \ell_\beta)$, or equivalently $R^Z_{\alpha \beta}$.
  We have addressed these observables within the framework of the ISS(3,3) in a recent study~\cite{Abada:2023raf}.

\section{Results}\label{sec:results}
We proceed to discuss the most important points of our analysis, numerically assessing the comparative prospects 
of observing cLFV $2\to2$ scatterings at $\mu$TRISTAN, in view of other cLFV constraints (arising from low-energy searches and $Z$ decays). As already mentioned, we will also evaluate angular distributions. Our study relies on several beam configurations (cf. Table~\ref{tab:muT:com}), and we do not take into account the possibility of beam polarisation.
Finally we will comment on the unique power of $\mu$TRISTAN to probe $\mu-\tau$ sector flavour violation.

We illustrate most of our results relying (for simplicity) on the minimal ad-hoc SM extension; we conclude our study by evaluating  the prospects of $\mu$TRISTAN regarding the ISS(3,3).

\subsection{General prospects: ``3+2'' ad-hoc extension}
To illustrate the prospects of this ``3+2'' extension, we consider the following benchmark choice of parameters: 
\begin{equation}
    m_4 = m_5\,,\quad s_{14} = s_{15} = 5\times 10^{-4}\,,\quad s_{24} = s_{25} = 0.01\,,\quad s_{34} = s_{35} = 0.1\,,\label{eqn:bench3p2}
\end{equation}
and further take several choices of mass of the heavy states, which  are assumed to be degenerate.
For simplicity, we begin by setting all phases to zero; we will subsequently consider their effect for specific observables. 

We begin by addressing the scattering  cross-sections, studying them for different lepton flavours in the final state, and as a function of the centre-of-mass energy.

\paragraph{Total cross-section}
Starting from the differential cross-section given in Eq.~\eqref{eqn:diffxsec}, we can integrate over the scattering angle in the centre-of-mass system to obtain the total cross-section as
\begin{equation}
    \sigma_\text{tot} = \int_{-1}^{+1}d\cos\theta \, \frac{d\sigma}{d\cos\theta}\,.
\end{equation}
For practical purposes however, one should keep in mind that detectors at a circular collider experiment do not typically exhibit a full ``$4\pi$-coverage''; consequently, one has to impose a cut-off of the integrated angle.
However, instead of imposing a cut on the angle in the laboratory or centre-of-mass frame, we opt to transform the integration variable to pseudo-rapidities via
\begin{eqnarray}
    \eta &=& -\log\left(\tan\left(\frac{\theta}{2}\right)\right)\,,\\
    d\cos\theta &=& \frac{1}{\cosh^2\eta}\, d\eta\,.\label{eqn:jac}
\end{eqnarray}
This has the advantage of allowing to impose some physically meaningful cuts on the phase space integration. 
We adopt here the value of rapidity coverage envisaged for the ATLAS HL-LHC upgrade, which is $|\eta|\leq 4$~\cite{ATLAS:2017svb}. 
After consistently rewriting the squared amplitude with $\eta$ rather than $\cos\theta$ and multiplying the cross-section with the Jacobian of Eq.~\eqref{eqn:jac}, the phase space integration thus becomes
\begin{equation}
    \sigma_\text{tot} = \int_{\eta_\text{min}}^{\eta_\text{max}} \frac{1}{\cosh^2\eta}\, d\eta\,  \frac{d\sigma}{d\eta}\,.
\end{equation}

\begin{figure}[h!]
    \centering
    \mbox{\includegraphics[width=0.48\linewidth]{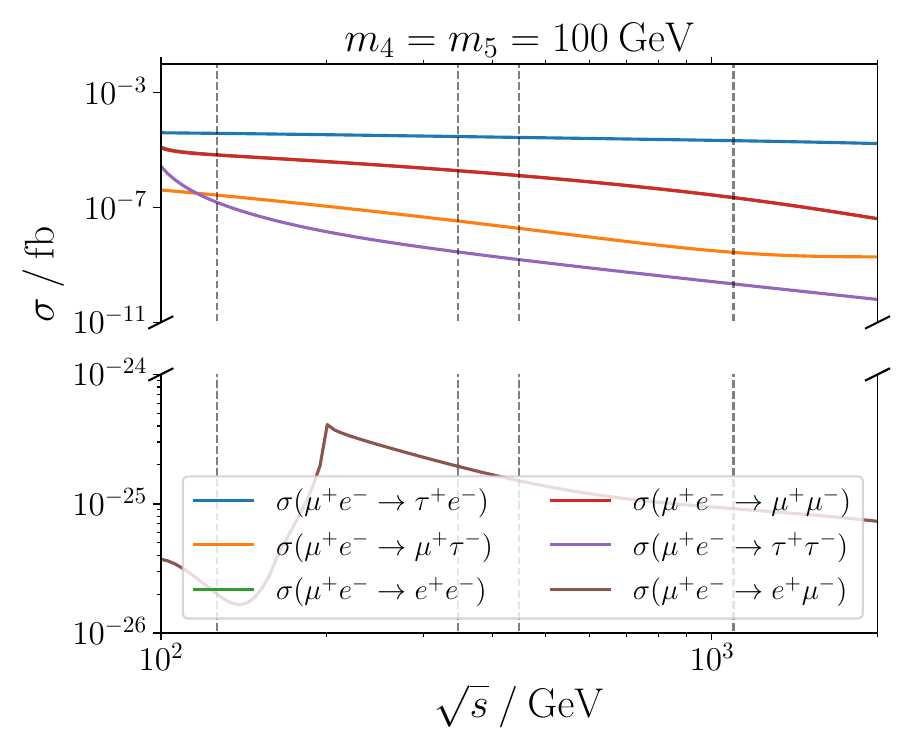}\includegraphics[width=0.48\linewidth]{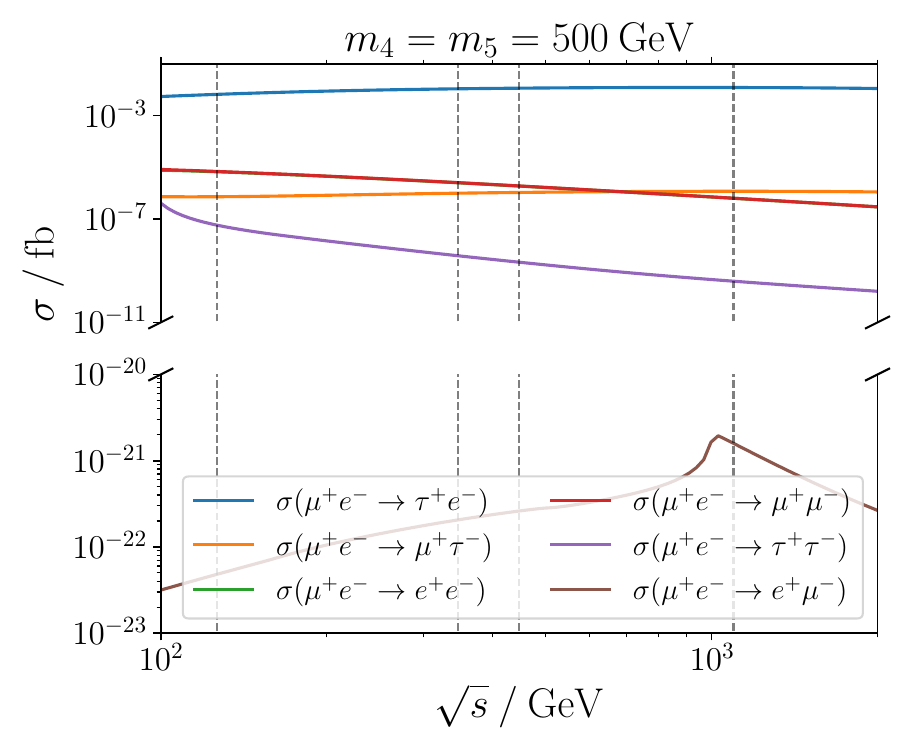}}\\
    \mbox{\includegraphics[width=0.48\linewidth]{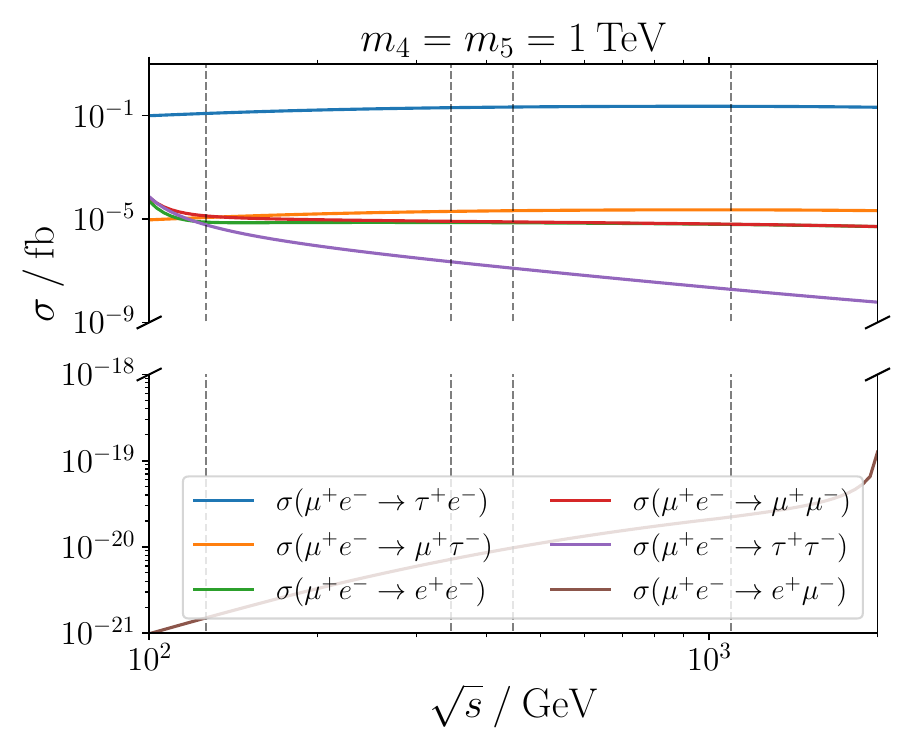}\includegraphics[width=0.48\linewidth]{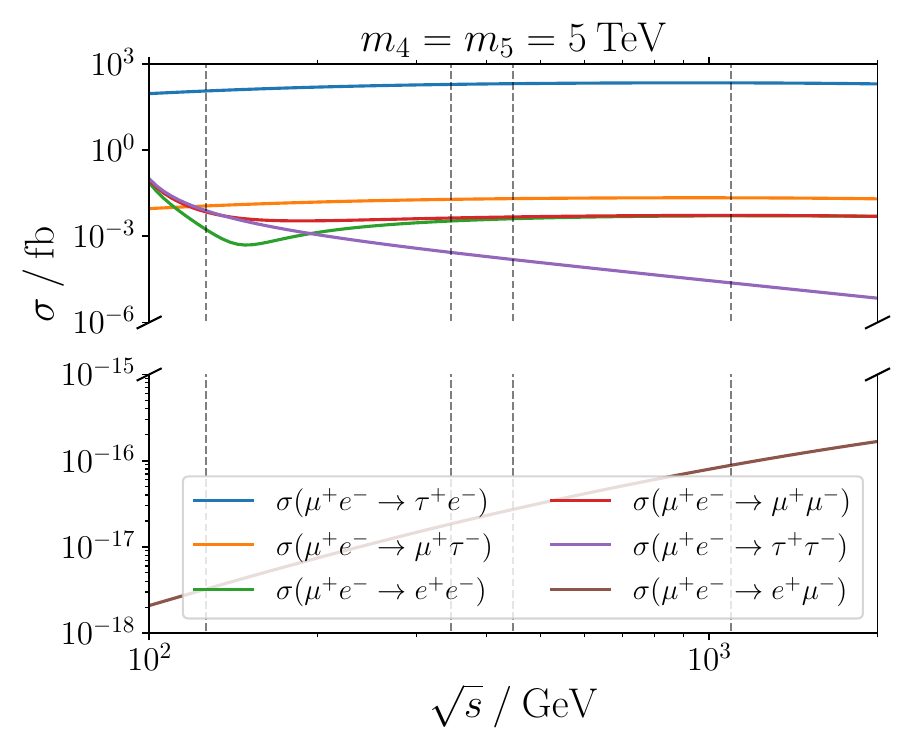}}
    \caption{Behaviour of the different cross-sections for $\mu^+ e^- \to \ell_\alpha^+ \ell_\beta^-$ with respect to $\sqrt{s}$. The mixing angles are set as given in Eq.~\eqref{eqn:bench3p2}, and we show  the results for four different choices of HNL masses with $m_4 = m_5 \in \{0.1, 0.5, 1, 5\}\:\mathrm{TeV}$ (from left to right from top to bottom). The dashed vertical lines denote the proposed centre-of-mass energies $\sqrt{s}\in \{126.5, 346.4, 447.2, 1095.4\}\:\mathrm{GeV}$.}
    \label{fig:baby3p2_vs_s}
\end{figure}

The results are displayed in Fig.~\ref{fig:baby3p2_vs_s}, 
in which we display the cross-sections for different flavour contents of the final state, with respect to 
$\sqrt s$, for different values of the heavy sterile states. In each panel we also include the proposed $\mu$TRISTAN centre-of-mass energies $\sqrt{s}\in \{126.5, 346.4, 447.2, 1095.4\}\:\mathrm{GeV}$ (cf. Table~\ref{tab:muT:com}). 
We notice that the benchmark point chosen in Eq.~\eqref{eqn:bench3p2} is mostly consistent with low-energy cLFV observables, as will be discussed later in greater detail.

As manifest, processes in association with the t-channel penguins 
(with either spectator-electron or spectator-muon fermion lines - see the first two diagrams of Fig.~\ref{fig:diagram:penguin} - such as $\mu^+e^- \to \tau^+e^- $ in blue and $\mu^+e^- \to \mu^+\tau^- $ in orange) exhibit a near independence with respect to $\sqrt s$ (only for much larger values of $\sqrt s$, lying outside the range here displayed, would there be a decrease in the -cross-section).
For $e^+e^-$ and $\mu^+\mu^-$ final states (the green and red lines respectively), there are competing s- and t-channel contributions: for lower regimes of $\sqrt s$, the s-channel contribution dominates (decreasing with $\sqrt s$); for intermediate c.o.m. energies, one can have interference between s- and t-channel contributions; ultimately for larger values of $\sqrt s$, t-channel contributions dominate, leading to an increase in the scattering cross-section. 
We highlight that this is only fully visible for the case of $m_{4,5} = 5$~TeV.
For final states composed of tau pairs (displayed in purple), and as can be seen from 
Fig.~\ref{fig:baby3p2_vs_s}, s-channel contributions are relevant for the explored regimes (c.o.m. energy and mass scale of the HNLs). 
Finally, one can verify that box contributions, shown in brown (as those at the source of 
$\mu^+ e^- \to e^+ \mu^-$) are in general subdominant. 

\begin{figure}[h!]
    \centering
    \mbox{\includegraphics[width=0.48\linewidth]{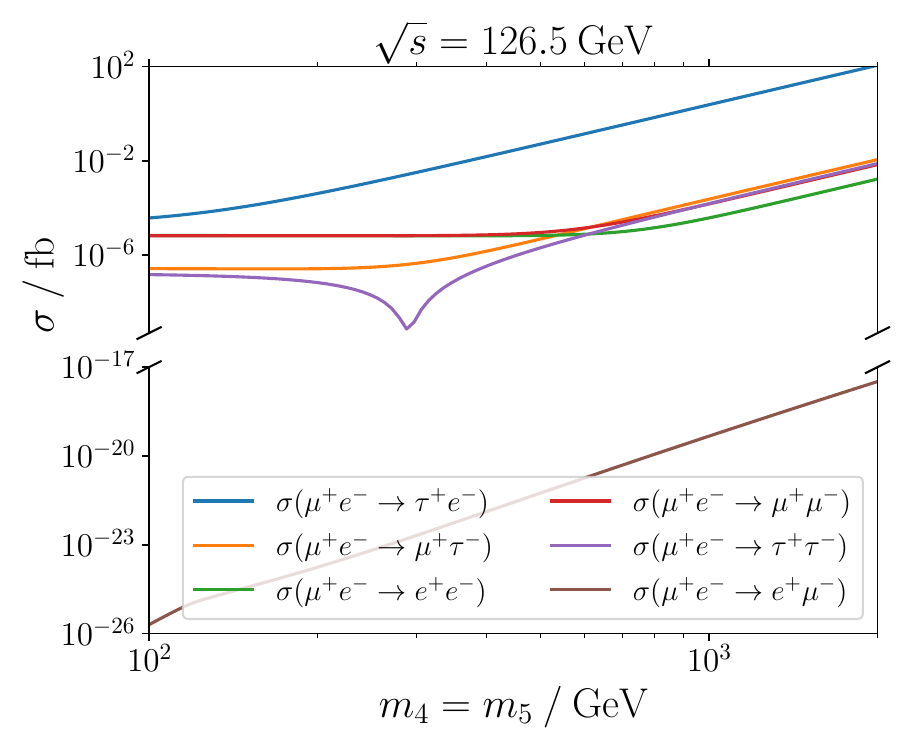}\includegraphics[width=0.48\linewidth]{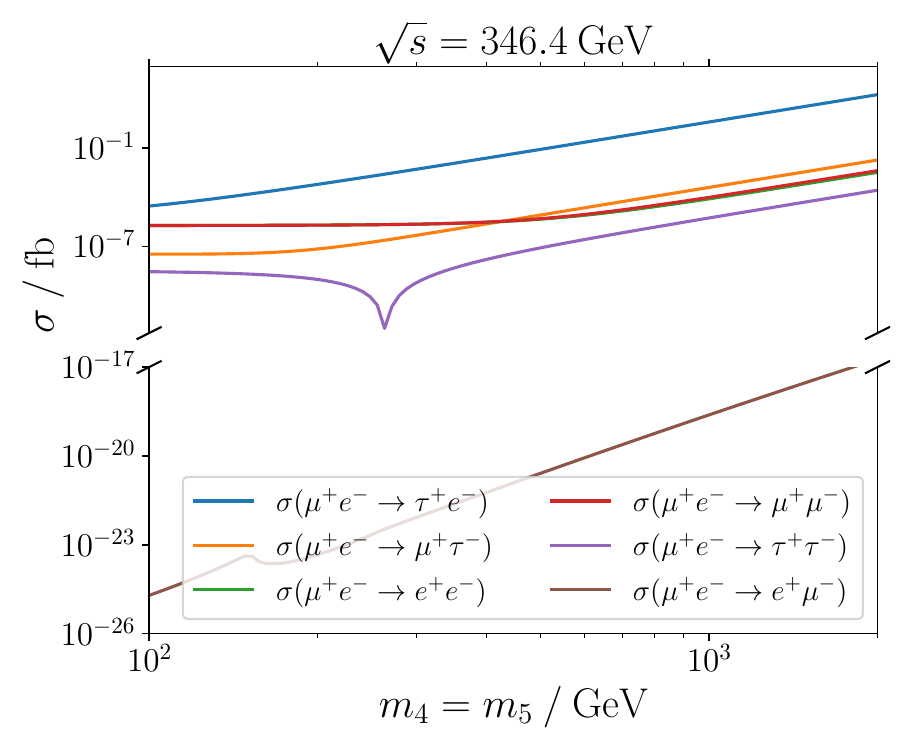}}\\
    \mbox{\includegraphics[width=0.48\linewidth]{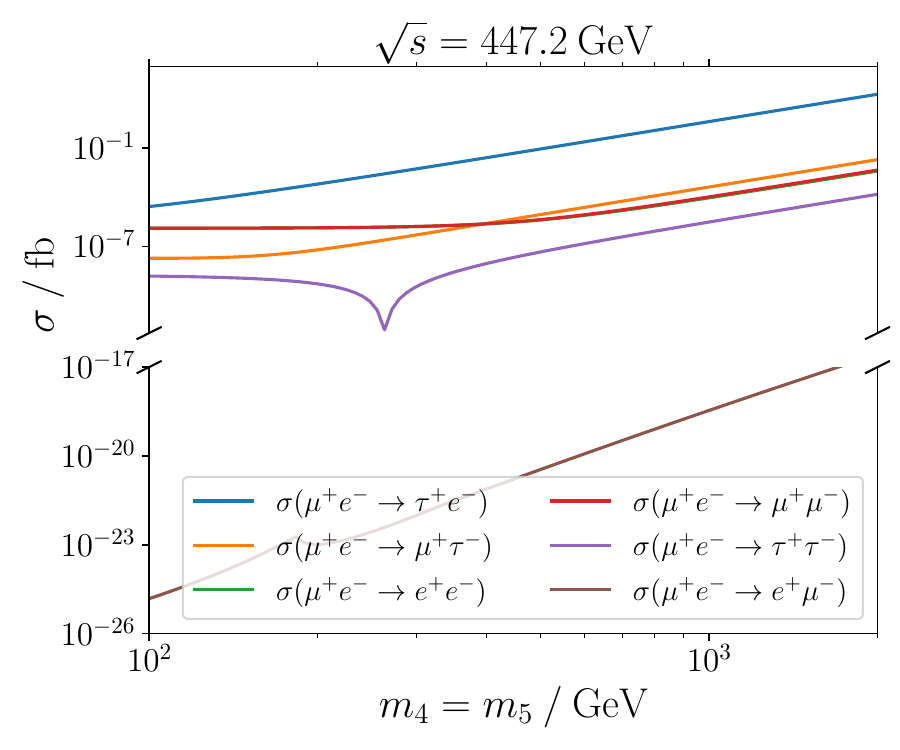}\includegraphics[width=0.48\linewidth]{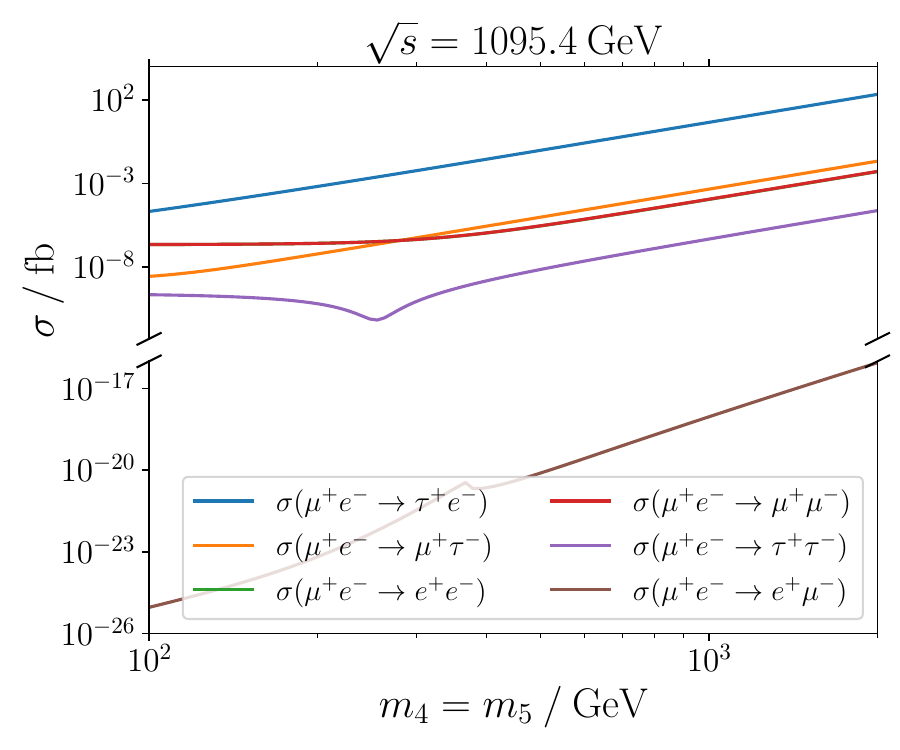}}
    \caption{Behaviour of the different cross-sections $\mu^+ e^- \to \ell_\alpha^+ \ell_\beta^-$ with respect to $m_4 = m_5$. The mixing angles are set as given in Eq.~\eqref{eqn:bench3p2}, and we show  the results for the four proposed centre-of-mass energies $\sqrt{s}\in \{126.5, 346.4, 447.2, 1095.4\}\:\mathrm{GeV}$ (from left to right from top to bottom).}
    \label{fig:baby3p2_vs_m4m5}
\end{figure}

A complementary view regarding the behaviour of the cross-section with respect to the mass scale of the heavy spectrum can be found in Figure~\ref{fig:baby3p2_vs_m4m5}. 
(Since in this ad-hoc realisation the HNL masses are varied while keeping couplings fixed, one trivially recovers the expected increase of the cross-section for increasing masses of the new heavy fermions.) 
The small ``dip'' occurring in the line corresponding to $\mu^+ e^- \to \tau^+ \tau^-$ reflects the threshold corresponding to the $W^+W^-$ mediated loop on the associated s-channel penguin contribution.

\begin{figure}[h!]
    \centering
    \mbox{\includegraphics[width=0.48\linewidth]{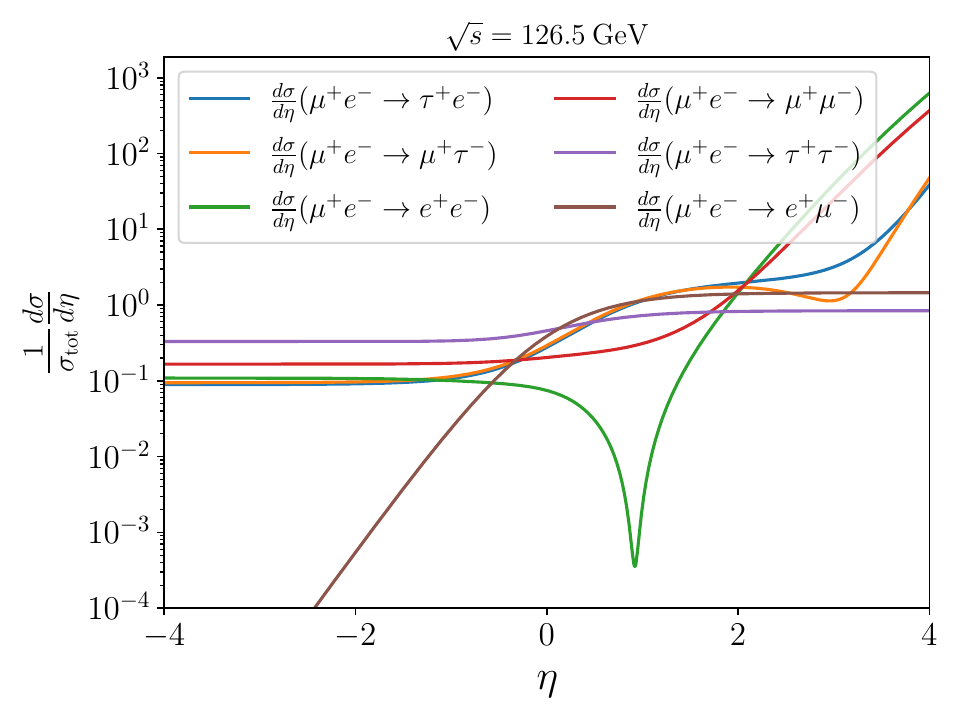}\includegraphics[width=0.48\linewidth]{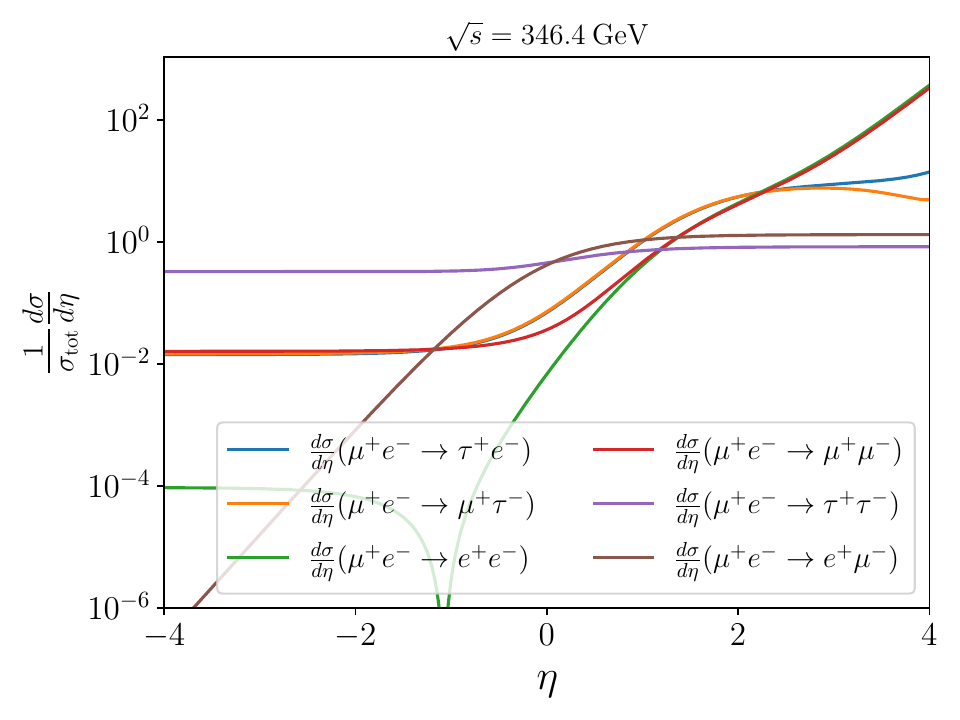}}\\
    \mbox{\includegraphics[width=0.48\linewidth]{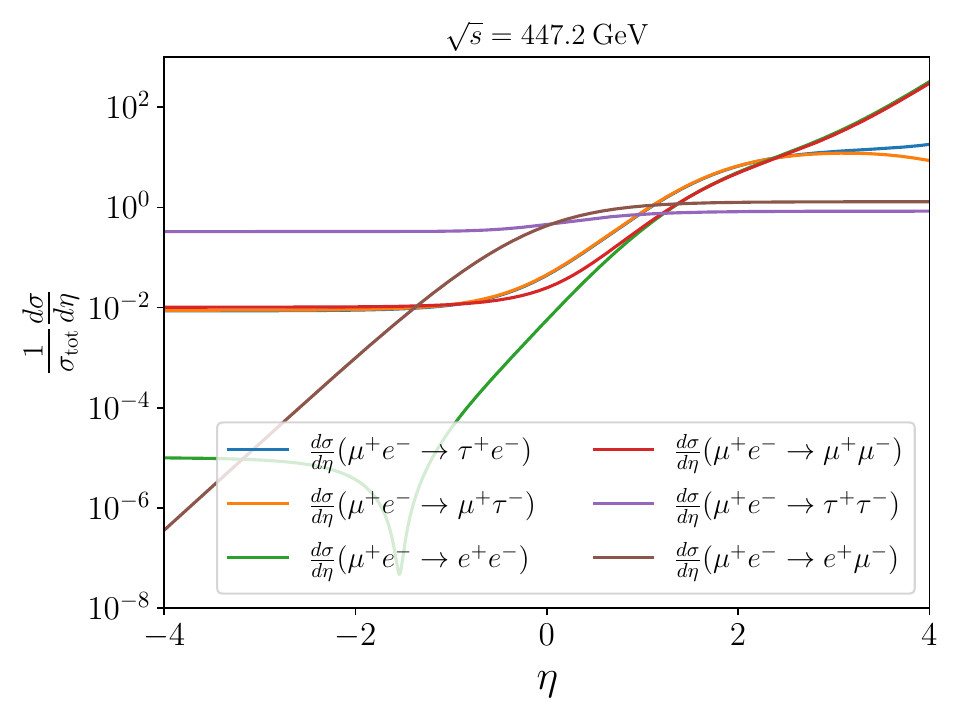}\includegraphics[width=0.48\linewidth]{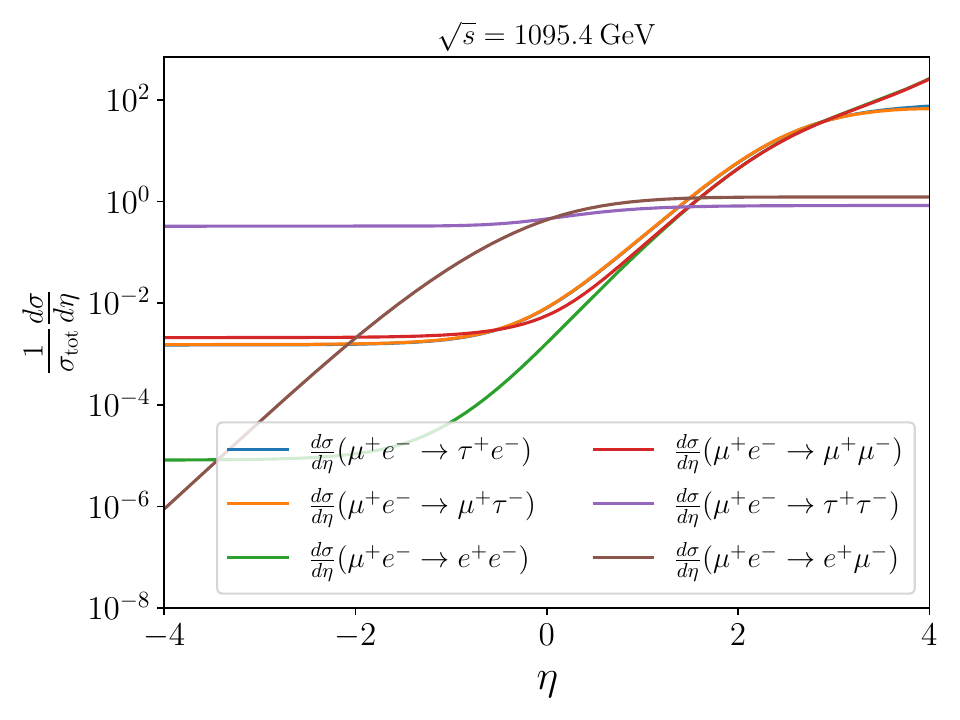}}
    \caption{Behaviour of the differential distribution 
    $d\sigma/d\eta$ (normalised to $\sigma_\text{tot}$ with respect to the pseudo-rapidity $\eta$, for the different processes $\mu^+ e^- \to \ell_\alpha^+ \ell_\beta^-$ under discussion. The mixing angles are as given in Eq.~\eqref{eqn:bench3p2}, and we have set $m_4 = m_5=1$~TeV. The panels correspond to the proposed centre-of-mass energies $\sqrt{s}\in \{126.5, 346.4, 447.2, 1095.4\}\:\mathrm{GeV}$ (from left to right from top to bottom).}
    \label{fig:baby3p2:dis_eta_vs_eta}
\end{figure}

\begin{figure}
    \centering
    \mbox{\includegraphics[width=0.48\linewidth]{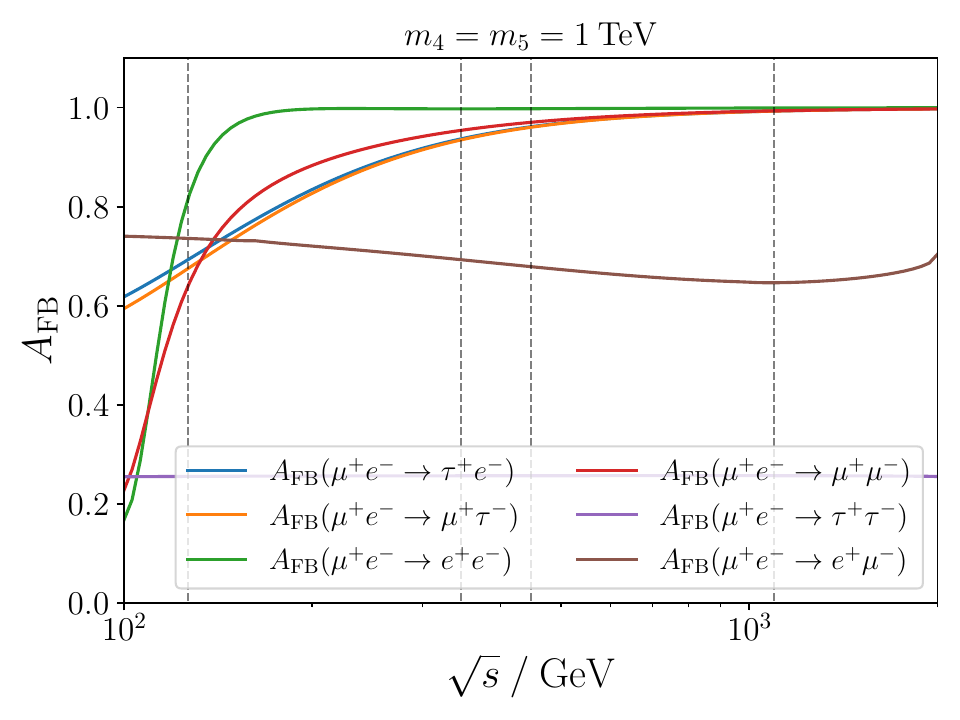}\includegraphics[width=0.48\linewidth]{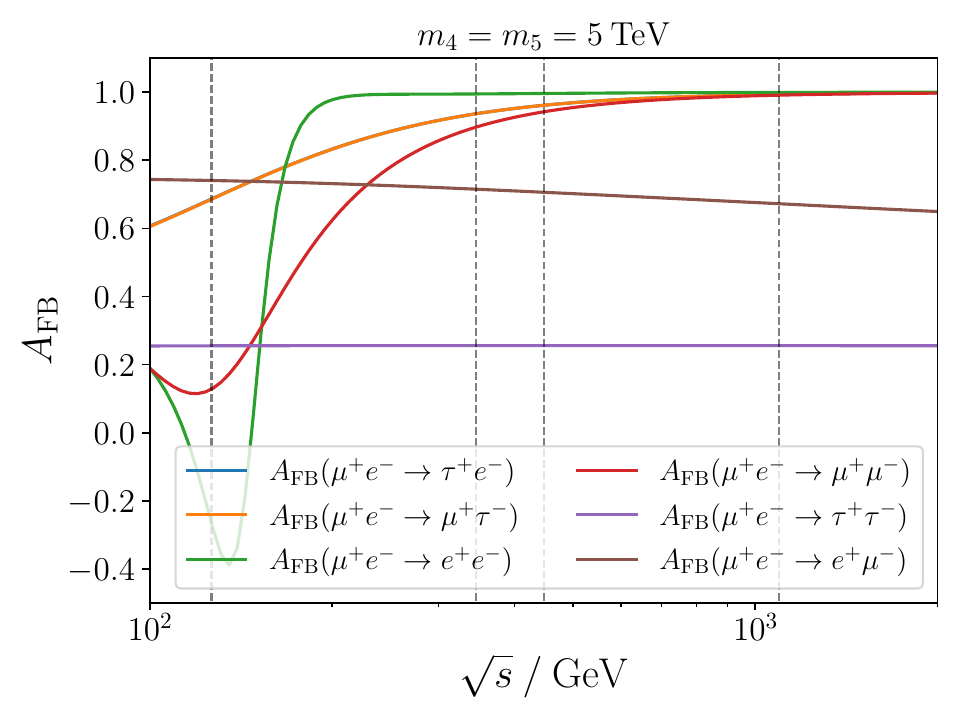}}
    \caption{Forward-backward asymmetry with respect to the c.o.m energy for two choices of the HNL masses. The mixing angles are as given in Eq.~\eqref{eqn:bench3p2}.
    }
    \label{fig:AFBvss}
\end{figure}

\begin{figure}
    \centering
    \mbox{\includegraphics[width=0.48\linewidth]{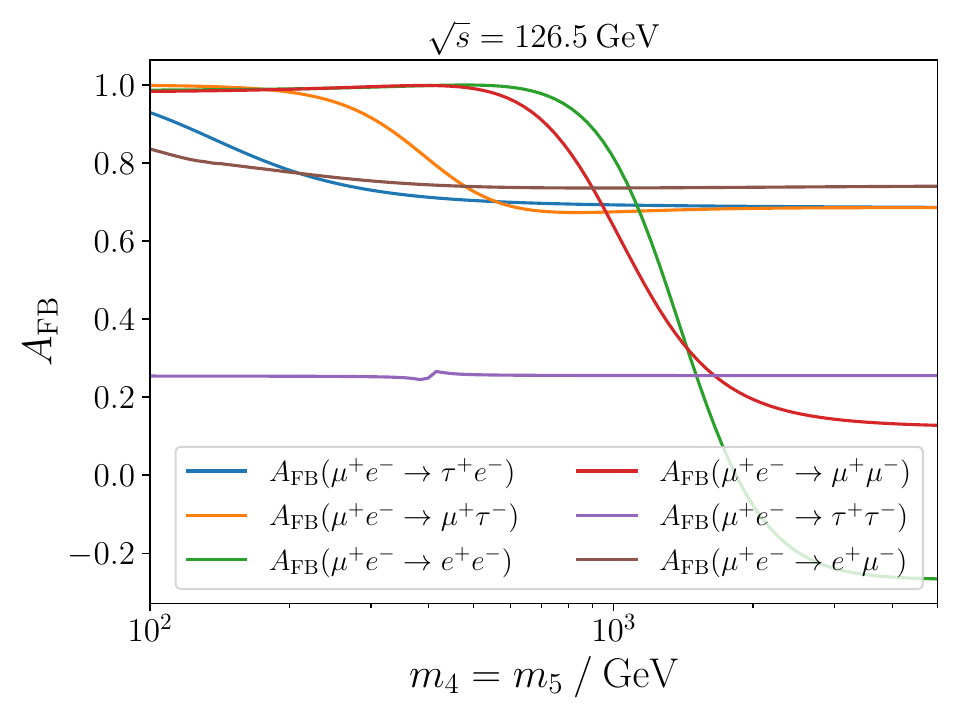}\includegraphics[width=0.48\linewidth]{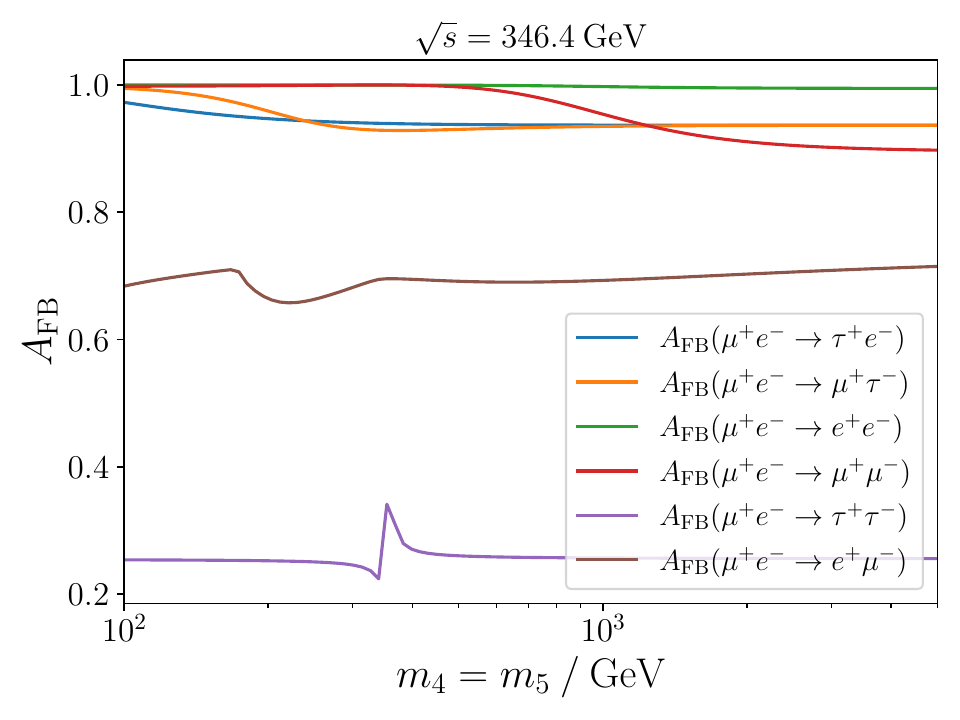}}
    \caption{Forward-backward asymmetry with respect to the HNL masses for two choices of the c.o.m energy. The mixing angles are as given in Eq.~\eqref{eqn:bench3p2}.}
    \label{fig:AFBvsm}
\end{figure}

\paragraph{Pseudo-rapidity distributions and forward-backward asymmetry}
It is also relevant to consider the prospects of these high-energy cLFV observables regarding angular distributions: in Figure~\ref{fig:baby3p2:dis_eta_vs_eta} we present an overview of 
differential distribution with respect to the pseudo-rapidity (normalised to the total cross-section, in order to improve the visibility of the results), setting $m_4 = m_5=1$~TeV. As manifest from all panels, processes exhibiting a strong t-channel penguin dominance are associated to a considerably larger number of events in the forward region ($\eta > 0$), which potentially allows for efficient signal and background discrimination.
In particular, this is the case of $\tau^+ e^-$, $\mu^+ e^-$, 
$e^+ e^-$ and $\mu^+ \mu^-$.
In Figure~\ref{fig:baby3p2:dis_eta_vs_eta} one can also see the different interference patterns between s- and t-channel penguins. This is especially true for the $\mu^+e^-\to e^+e^-$ and $\mu^+e^-\to \mu^+\mu^-$ scatterings. 
While the $\mu^+e^-\to e^+e^-$ distribution displays a cancellation (or rather a ``dip'') for most of the considered $\sqrt{s}$, this feature is absent in the $\mu^+e^-\to \mu^+\mu^-$, since the t-channel penguin enters differently in the amplitudes of Eq.~\eqref{eqn:general_amps} (see also Eqs.~(\ref{eqn:amppair1}-\ref{eqn:amppair2})).
Furthermore, we have considered the forward-backward asymmetry, $A_\text{FB} (\mu^+ e^- \to \ell_\alpha^+ \ell_\beta^-)$.
For $2\to2$ processes, the forward-backward asymmetry is commonly defined as 
\begin{eqnarray}
    A_\text{FB} &=& \frac{\sigma_\text{F} - \sigma_\text{B}}{\sigma_\text{F} + \sigma_\text{B}}\,,\\
    \sigma_\text{F} = \int_0^{\eta_\text{max}} d\eta \frac{d\sigma}{d\eta}\,,&\phantom{=}&\sigma_\text{B} = \int_{\eta_\text{min}}^{0} d\eta \frac{d\sigma}{d\eta}\,,
\end{eqnarray}
in which, as previously stated, we have imposed the cut $|\eta|\leq 4$ for the computation of $\sigma_\text{F,B}$. 
We present some numerical results on $A_{\text{FB}}$: first, as a function of the c.o.m energy in Figure~\ref{fig:AFBvss} (for two fixed values of the heavy fermion masses, 1~TeV and 5~TeV); versus the (degenerate) masses of the HNL (for two of the proposed c.o.m. energies, $\sqrt s=126.5$~GeV and $\sqrt s=346.4$~GeV), as shown in Figure~\ref{fig:AFBvsm}.
As can be seen, for the two nominal operating $\sqrt{s}$ values, a measurement of the different forward-backward asymmetries could provide complementary information to that solely arising from the total cross-section; as visible from the illustrative examples presented, $A_{\text{FB}}$ is clearly sensitive to the topology of the cLFV process, and can help clarifying the underlying nature of the (dominant) contribution to the cLFV scatterings.
Moreover, a combination of both measurements could hint at the mass-scale of the HNL at the source of the  cLFV transitions.

\begin{figure}
    \centering
    \mbox{\includegraphics[width=0.48\linewidth]{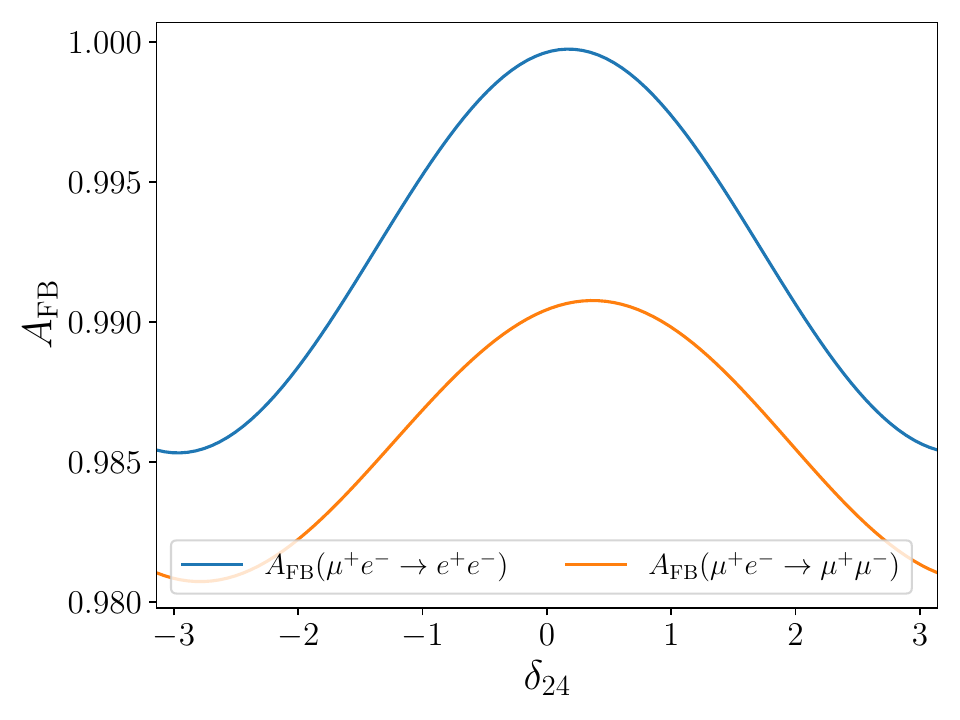}\includegraphics[width=0.48\textwidth]{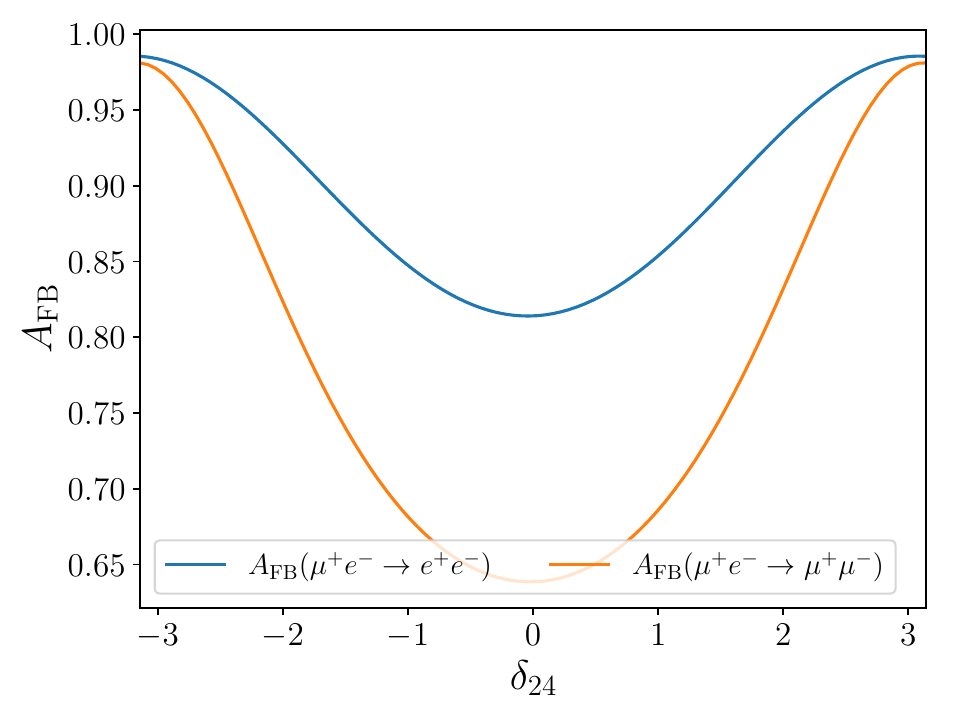}}
    \caption{Variation of the forward-backward asymmetries for $ee$ and $\mu\mu$ final states with respect to the CP phase $\delta_{24}$, with the mixing angles as given in Eq.~\eqref{eqn:bench3p2}. We have set $\sqrt{s}=126.5\:\mathrm{GeV}$. On the left, $m_4 = m_5 = 500\:\mathrm{GeV}$, while on the right $m_4 = m_5 = 1\:\mathrm{TeV}$.}
    \label{fig:AFB_vs_phase}
\end{figure}

Furthermore, the different $A_\mathrm{FB}$ are in principle sensitive to new CP-violating phases in the generalised lepton mixing matrix. In order to illustrate this, we have considered the effect of a (Dirac) phase, 
$\delta_{24}$, and the results are shown in Figure~\ref{fig:AFB_vs_phase}.
This can be simply understood by considering that the phases lead to a suppression of the $Z$-penguins (depending on the flavour content of the transitions), such that the interference between the s- and t-channel penguins in $\mu^+e^-\to e^+ e^-$ and $\mu^+e^-\to \mu^+ \mu^-$ will be altered by the presence of non-vanishing CP phases. 

\paragraph{Complementary approaches to cLFV: low energies and $\mu$TRISTAN}

The results so far obtain suggest that $\mu$TRISTAN offers very interesting possibilities to search for cLFV 
$2\to2$ scatterings, as those induced by the presence of HNL states (and a non-unitary leptonic mixings).
The question that must now be addressed is how competitive the $\mu$TRISTAN cLFV searches are when compared to the sensitivity of low-energy searches and those performed at the $Z$-pole at FCC-ee.

\begin{figure}[h!]
    \centering
    \mbox{\includegraphics[width=0.48\linewidth]{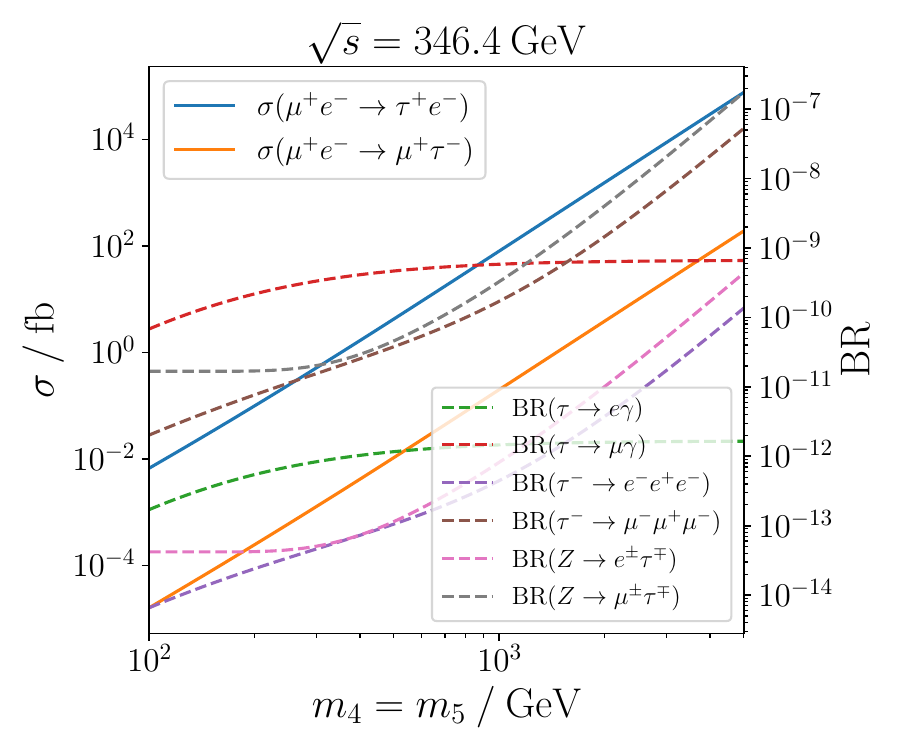}\includegraphics[width=0.48\linewidth]{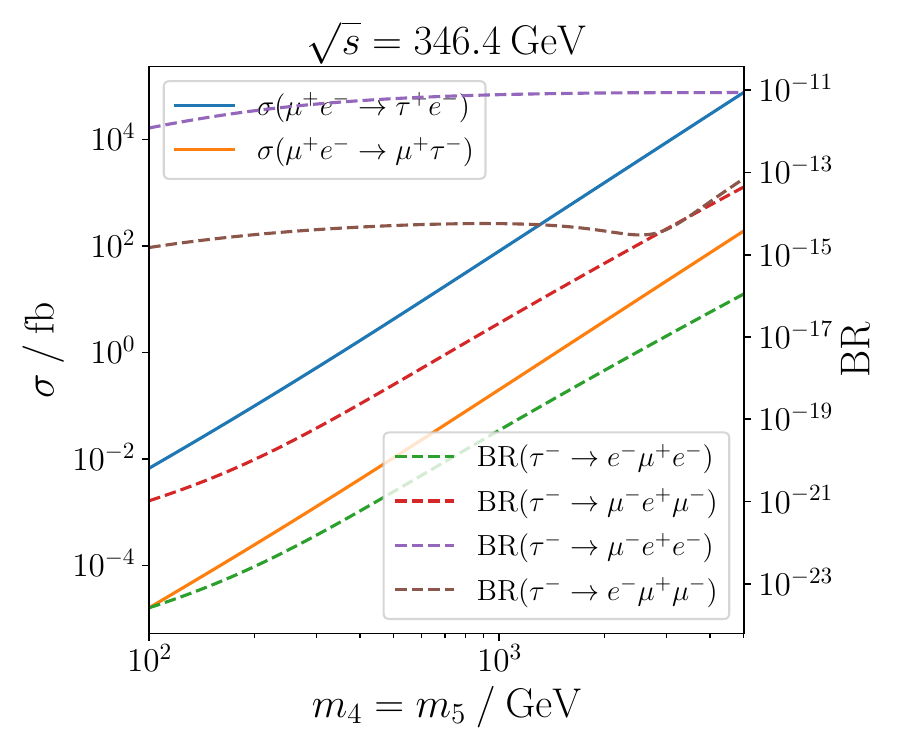}}\\
    \mbox{\includegraphics[width=0.48\linewidth]{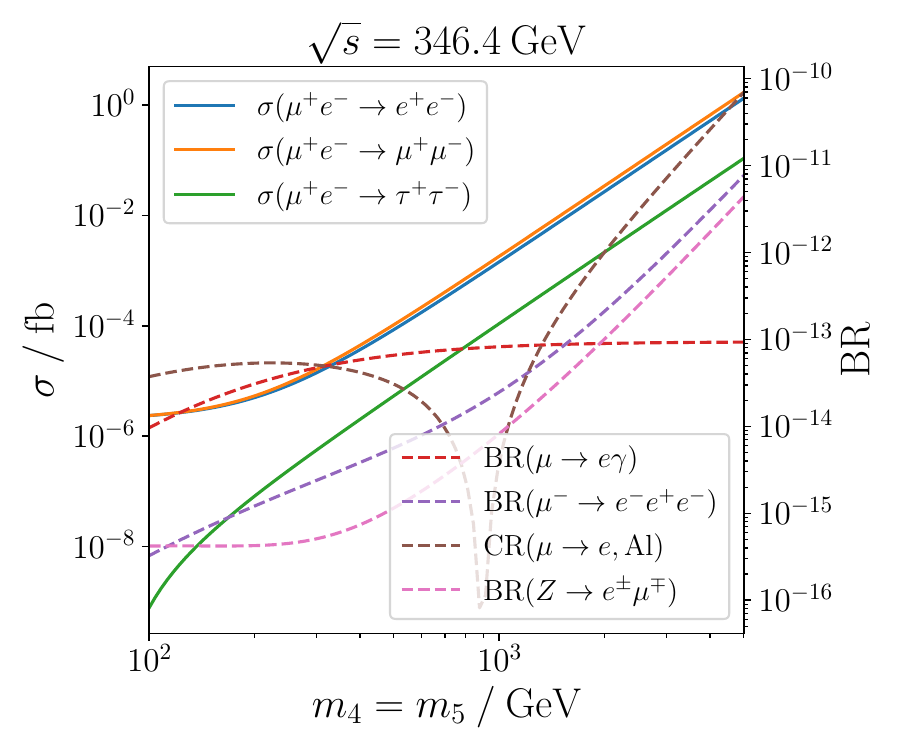}\includegraphics[width=0.48\linewidth]{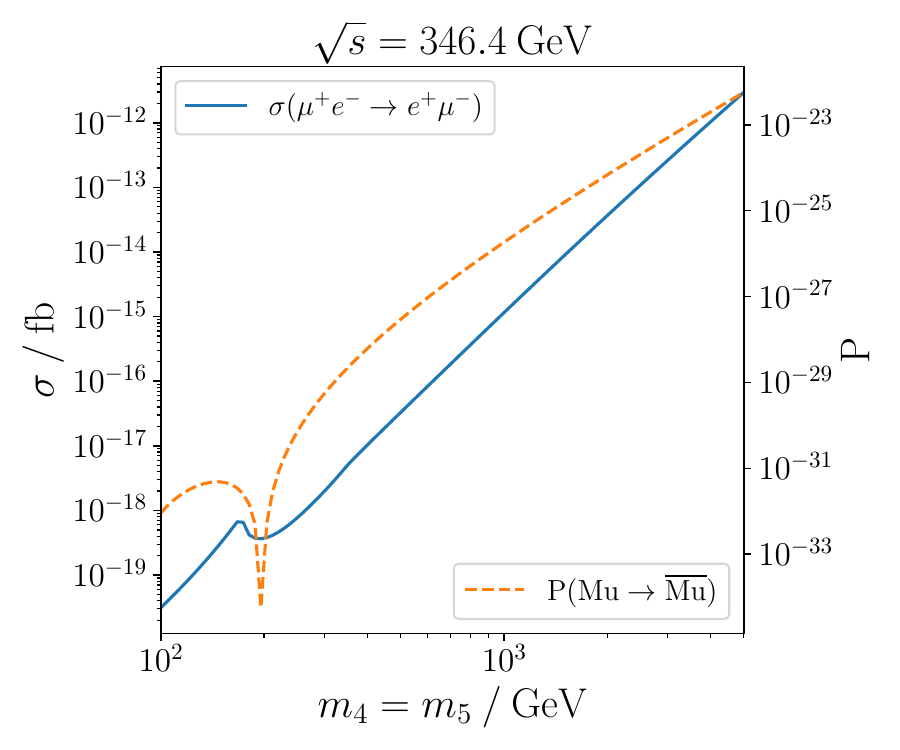}}
    \caption{Comparative prospects for cLFV processes: $\sigma(\mu^+ e^- \to \ell_\alpha^+ \ell_\beta^-)$ as a function of the (degenerate) HNL masses for $\sqrt s=346.4$~GeV. On the secondary $y$-axes (right hand-side of each plot) we display the rates for associated cLFV processes: radiative and 3-body decays, neutrinoless muon-conversion in matter, muonium oscillations and cLFV $Z$ boson decays.  }
    \label{fig:sig_vs_cLFV}
\end{figure}

For the illustrative flavour-mixing benchmark here considered (in the context of the minimal ad-hoc extension), it is interesting to notice that $\mu$TRISTAN offers indeed very good prospects to discover HNL through cLFV $\mu^+ e^- \to \ell_\alpha^+ \ell_\beta^-$ scatterings. 
In order to substantiate this, we compare the scattering cross-sections $\mu^+ e^- \to \ell_\alpha^+ \ell_\beta^-$ with cLFV processes involving $\mu-\ell_\alpha$ and/or 
$e-\ell_\beta$ transitions: for example, it is pertinent to compare $\mu^+ e^- \to \tau^+ e^-$ with $\tau \to 3\mu$ and  $Z\to \mu \tau$ decays.
This is carried out in Figure~\ref{fig:sig_vs_cLFV}, whose panels  offer a direct comparison of 
$\sigma(\mu^+ e^- \to \ell_\alpha^+ \ell_\beta^-)$ with rare cLFV  decays relying on the same flavour violating transitions. Taking $\sqrt s=346.4$~GeV, 
we compare the scattering cross-sections with the relevant branching ratios for varying values of the heavy sterile states. 

The comparison of $\sigma(\mu^+ e^- \to \tau^+ e^-)$ with the prospects for leptonic cLFV tau decays and $Z\to \ell \tau$ decays
(displayed in the  top-left panel of Fig.~\ref{fig:sig_vs_cLFV}) clearly shows that, 
for the ``3+2'' flavour benchmark under consideration, one can expect tens (or even hundreds) of thousands of events at $\mu$TRISTAN\footnote{Relying on a conservative estimate of the efficiency, $\mathcal{O}(1\%)$, and for the nominal luminosities, see Section~\ref{sec:muT}. Our estimate for the efficiency is further outlined below.}.
Notice that, apart from the muon observables ($\mu\to e\gamma$, $\mu\to3e$ and $\mu-e$ conversion), all predicted cLFV rates lie (mostly) beyond future observability (cf. Table~\ref{tab:cLFV_lep}), while the cLFV cross-sections are sizeable for the same considered masses (especially for $m_{4,5}\lesssim 1\:\mathrm{TeV}$).
The plots in this figure highlight how such an asymmetric lepton collider can play a competitive role in what concerns the observation of cLFV processes involving tau leptons.


It is also interesting to comment on a cLFV low-energy observable that strongly resembles $\mu^+ e^- \to e^+ \mu^-$ scattering: muonium oscillations. As can be seen from the lower right-hand panel of Fig.~\ref{fig:sig_vs_cLFV}, both the scattering processes at $\mu$TRISTAN, and the oscillation probability clearly lie beyond any prospect for observation.

We emphasise here the comparative study presented in Figure~\ref{fig:sig_vs_cLFV} should not be perceived as a generic feature of these minimal ad-hoc extensions. Nevertheless, it can be interpreted as a  {\it proof of merit} that $\mu$TRISTAN offers rich prospects for the observation of cLFV
- even in the absence of other signals ($Z$ decays and/or low-energy transitions).

\subsection{ISS(3,3)}
After a first approach to the cLFV observables that can be studied at $\mu$TRISTAN, relying on the simple minimal SM extension via 2 HNL, we now proceed to consider a complete model of neutrino mass generation. Contrary to the ad-hoc model, we recall that in such a case NP mass scales and couplings are intrinsically connected by the need to accommodate neutrino oscillation data. 
It is nevertheless important to notice that the behaviours discussed above for the ad-hoc minimal extension are representative of generic SM extensions via HNLs; the (relative) weights of the cLFV form-factors naturally reflect the actual model under consideration, but the qualitative features are expected to hold for the case of a complete model as the (3,3) realisation of the ISS (described in Section~\ref{sec:HNL:model:constraints}).

In particular, the goal of this final study is to highlight  
through a complete NP model the complementary role that  $\mu$TRISTAN can play in the landscape of cLFV searches. 

In order to illustrate our findings in a concrete and efficient manner, we rely on the modified Casas-Ibarra parametrisation (see Eq.~\eqref{eqn:ISSCI}) to determine the Yukawa couplings while in agreement with oscillation data. 
We set the orthogonal Casas-Ibarra $R$-matrix to unity and assume a diagonal and universal $\mu_X=10^{-6}\:\mathrm{GeV}$ (at this first stage); we further assume $M_R$ to be diagonal and universal. Concerning the active neutrino parameters, the PMNS entries have been set to the central values according to NuFit6.0~\cite{Esteban:2024eli}, and the lightest neutrino mass to $m_0 = 10^{-5}\:\mathrm{eV}$ (we only consider normal ordering (NO)). 
The benchmark ``flavour-space'' thus obtained - since $M_R$ is not fixed - reflects an ensemble of realistic cases which have not yet been excluded by current bounds on cLFV processes nor from leading to large deviations from unitarity of the would-be PMNS.
\begin{figure}[h!]
    \centering
    \mbox{\includegraphics[width=0.48\linewidth]{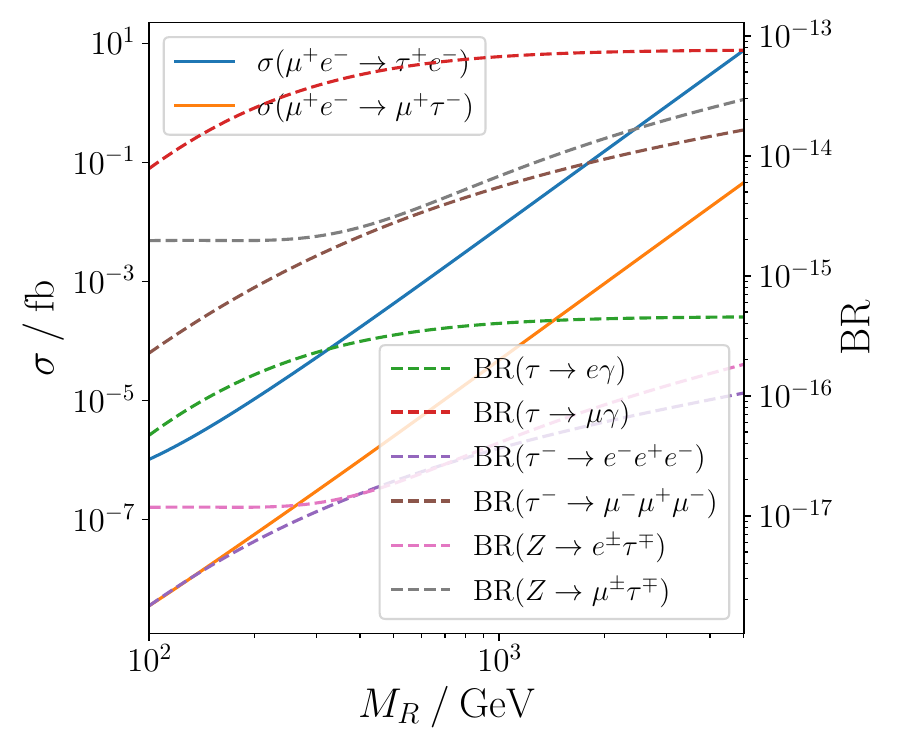}\includegraphics[width=0.48\linewidth]{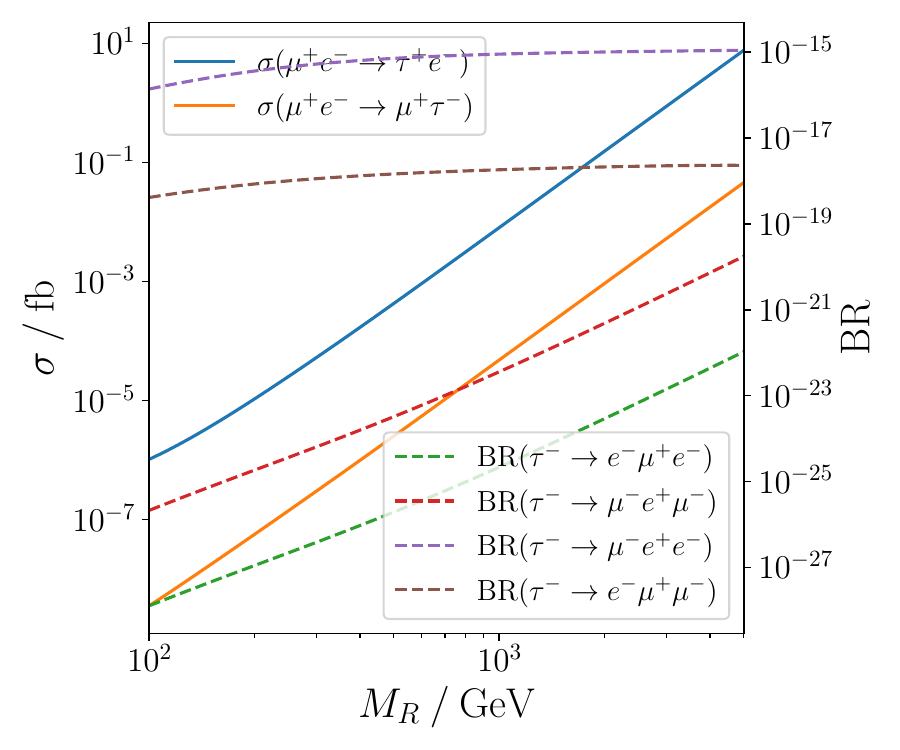}}\\
    \mbox{\includegraphics[width=0.48\linewidth]{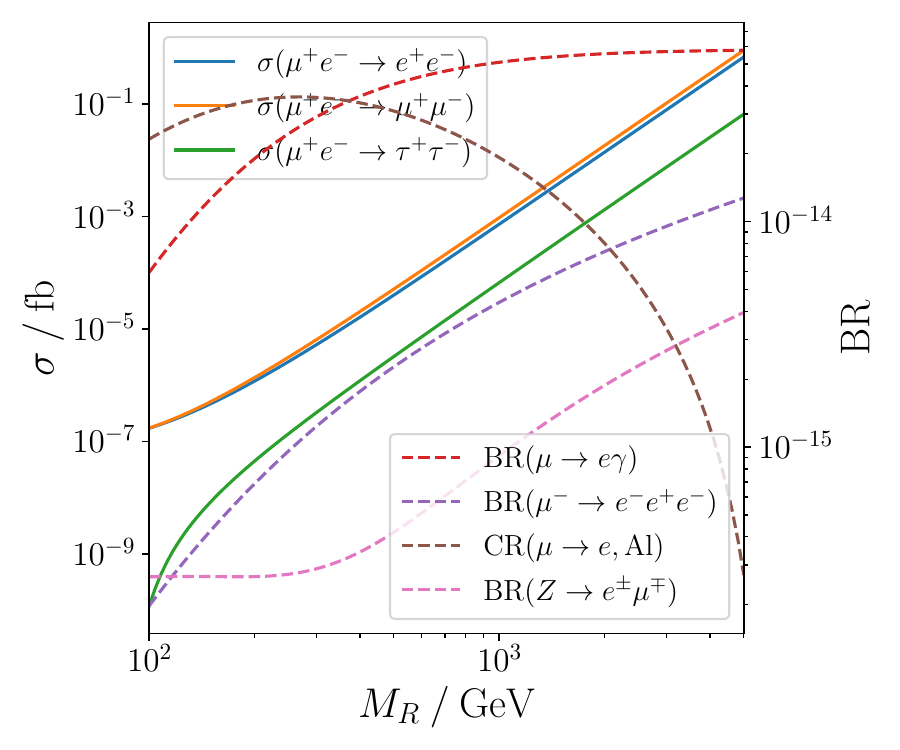}\includegraphics[width=0.48\linewidth]{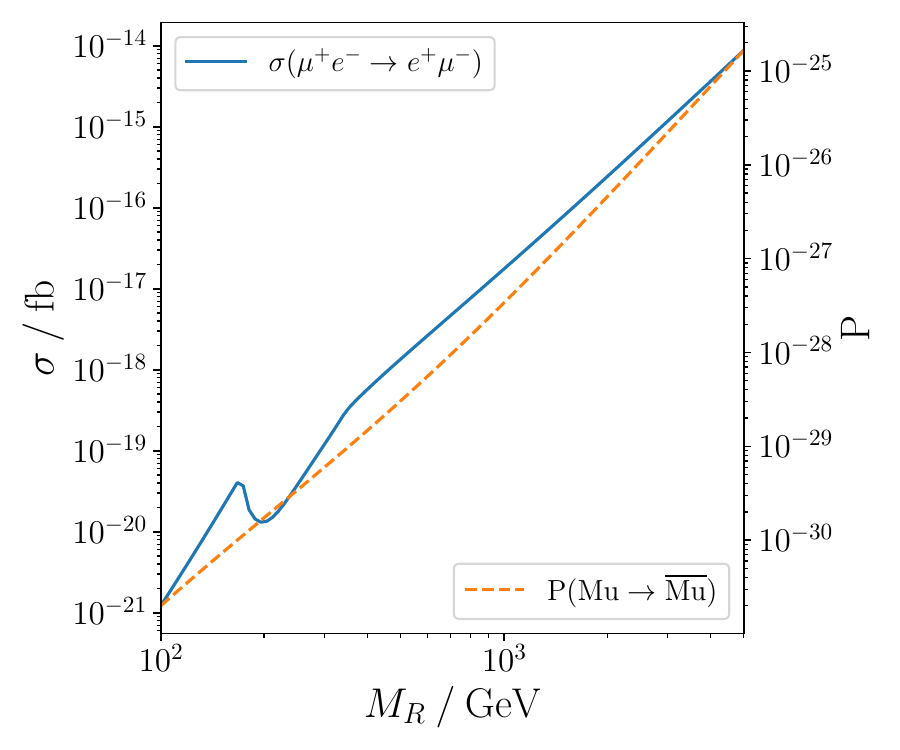}}
    \caption{Comparative prospects for cLFV processes: $\sigma(\mu^+ e^- \to \ell_\alpha^+ \ell_\beta^-)$ as a function of the (degenerate) $M_R$ masses, for $\mu_X = 10^{-6}\:\mathrm{GeV}$ and  $\sqrt s=346.4$~GeV. On the secondary $y$-axes (right side of each plot) we display the rates for associated cLFV processes: radiative and 3-body decays, neutrinoless muon-conversion in matter, muonium oscillations and cLFV $Z$ boson decays.}
    \label{fig:ISS_CI_cLFV}
\end{figure}

To begin our discussion, we reproduce in Figure~\ref{fig:ISS_CI_cLFV} an analogous study to that of 
Figure~\ref{fig:sig_vs_cLFV} (for the ad-hoc model), comparing $\sigma(\mu^+ e^- \to \ell_\alpha^+ \ell_\beta^-)$ with several cLFV branching rates, as a function of the (degenerate) $M_R$ masses.
As can be seen, the cLFV rates for processes involving $\tau$-leptons at low energies and in $Z$-decays all lie below any future sensitivities, while the cross-sections for $\mu^+e^-\to\tau^+e^-$ and $\mu^+e^-\to \mu^+\tau^-$ scattering can reach sizeable values, leading to hundreds or even thousands of events that could be potentially observed by $\mu$TRISTAN.
For $\mu-e$ violating processes, searches at $\mu$TRISTAN cannot compete with low-energy probes, in agreement with the expectations arising from the study of the ad-hoc model of the previous subsection.

\bigskip
In view of the sizeable cross-sections, it is worthwhile to carry out a na\"ive analysis of the potential of the 
$\mu$TRISTAN collider in constraining the parameter space of the ISS(3,3).
Still relying on the  Casas-Ibarra parametrisation - as described before -  we nevertheless depart from the simple hypothesis of fixed $\mu_X$, and explore the ISS(3,3) space spanned by the NP scale $M_R$ and the size of the LNV parameter, $\mu_X$.  

In the absence of concrete detector designs, a full study of possible backgrounds and systematics is clearly beyond the scope of this work.
However, we can make some preliminary estimates and identify sources of the latter.
Firstly, a foreseeable improvement from the theory side is the inclusion of QED corrections (virtual and real) which will affect not only the overall cross-sections but also the event shapes.
A proper computation of the $2\to3$ scattering cross-sections (with an additional photon from initial state or final state radiation) would also allow scanning over ``effective'' $\sqrt{s}$ c.o.m. energies (making use of the so-called radiative return), similarly to what was done at LEP.

Experimentally, and if we assume the PMNS paradigm,
one can envisage backgrounds from $\mu^+ e^-\to\ell_\alpha^+\ell_\beta^-\nu\nu$ production.
As an example, we estimate the total cross-section of $\mu^+e^-\to\tau^+e^-\nu\nu$ background to be (at leading order) $\sigma_\text{bkg}\simeq 21\:\mathrm{fb}$.
This can be significantly reduced by imposing very basic cuts by selecting events with small (or vanishing) missing energy $\slashed{E}_T \leq 10\:\mathrm{GeV}$ and events that mostly lie in the forward region (see Figure~\ref{fig:baby3p2:dis_eta_vs_eta}).
By further demanding $2.0\leq\eta_\ell\leq 4.0$, we reduce the background cross-section to $\sigma_\text{bkg, cut}\simeq 6\times 10^{-4}\:\mathrm{fb}$, so that for an integrated luminosity of $1\:\mathrm{ab}^{-1}$
less than one background event is expected.
The signal efficiency for these very basic cuts is $25\%$;  combined with a conservatively estimated $\tau$-tagging efficiency of $40\%$~\cite{ATLAS:2024tzc} we can estimate the total signal efficiency at $\simeq 10\%$. 
Allowing for further signal loss due to charge and particle misidentification, as well as beam contamination due to in-flight decaying muons, we take a total signal efficiency of $1\%$ as a conservative estimate (while reducing the backgrounds to negligible levels).
We demand a detection of (at least) 10 signal events with negligible backgrounds for our sensitivity estimates, leading to a total significance of 
\begin{eqnarray}
    \mathcal S = \frac{S}{\sqrt{S + B}}\sim 3\,\sigma\,.
\end{eqnarray}

\begin{figure}
    \centering
    \mbox{\includegraphics[width=0.48\linewidth]{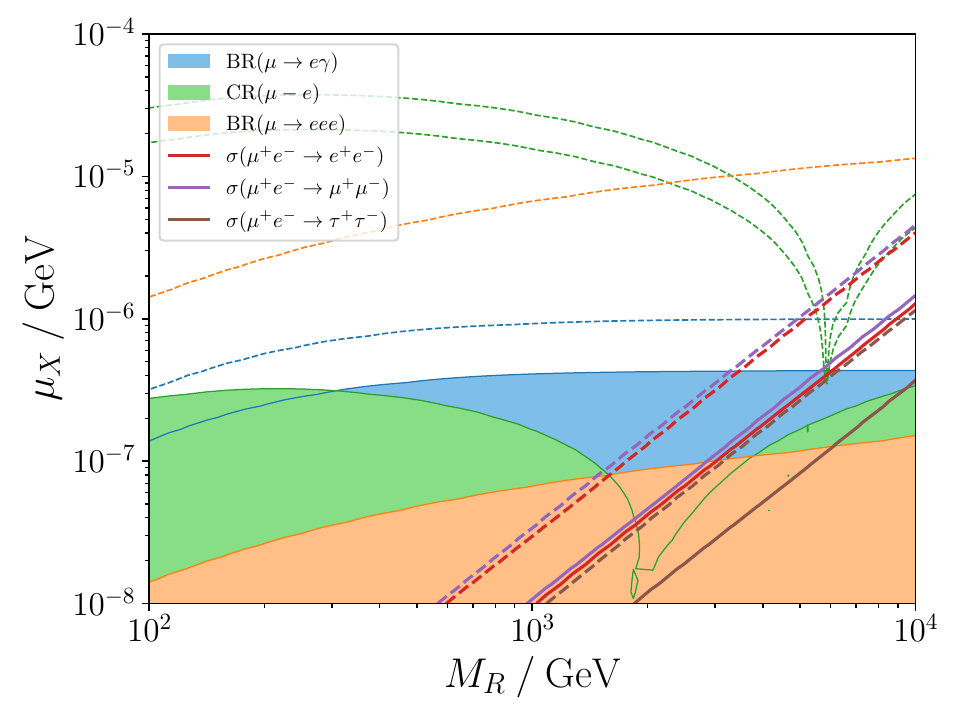}\includegraphics[width=0.48\linewidth]{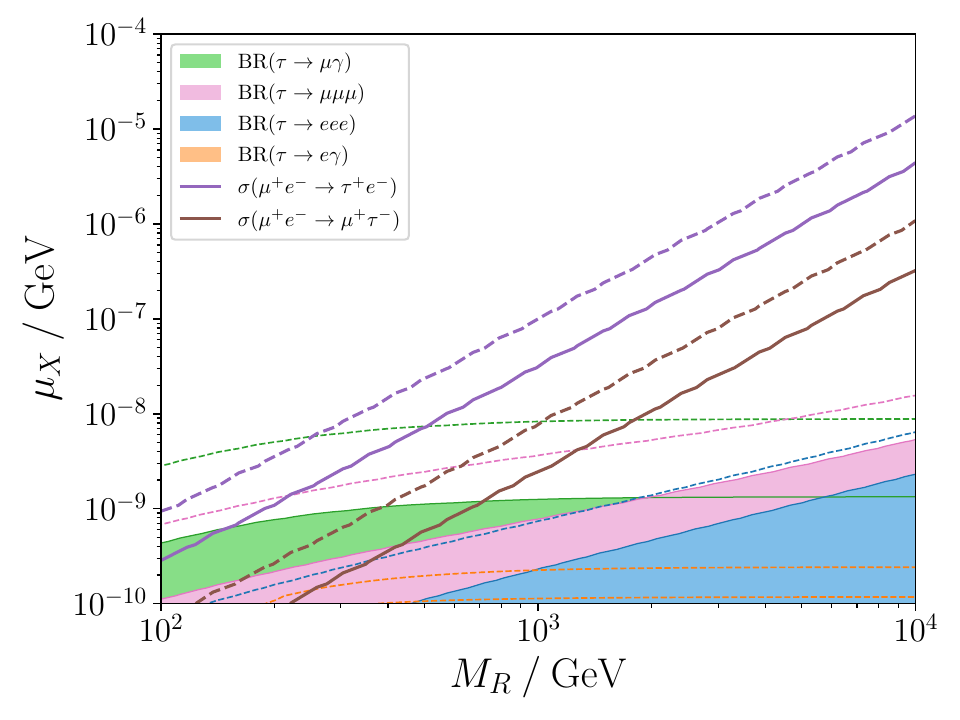}}
    \caption{Prospects for ISS(3,3) cLFV searches depicted in the $(M_R-\mu_X)$ plane:
    for bounds on low-energy cLFV processes, the filled contours correspond to current bounds and the dashed lines to (distinct) future sensitivities, see Table~\ref{tab:cLFV_lep}. (Notice that for $\mu-e$ conversion there are two dashed lines for the two future Mu2e and COMET sensitivities.)
    Expected $\mu$TRISTAN cross-sections for $1\%$ efficiency and $100\:\mathrm{fb}^{-1}$ luminosity ($1\:\mathrm{ab}^{-1}$) are given by solid (dashed) lines.}
    \label{fig:contour_CI}
\end{figure}

Our results are shown in Figure~\ref{fig:contour_CI} for $\mu-e$ flavour violation (on the left) and for  processes involving $\tau$-leptons (on the right).
We draw sensitivity contours in the $(M_R-\mu_X)$ plane for current bounds on low-energy cLFV processes (see legends) and future sensitivities of dedicated facilities (cf. Table~\ref{tab:cLFV_lep}), the latter corresponding to the dashed lines in the same colour as the filled contours.
Onto this we superimpose the sensitivities of a $\mu$TRISTAN collider obtained with the previously outlined analysis, 
assuming $100\:\mathrm{fb}^{-1}$ (solid lines) and $1\:\mathrm{ab}^{-1}$ (dashed lines) luminosities.

As can be seen, for $\mu-e$ sector flavour violation, searches at $\mu$TRISTAN cannot compete with low-energy processes at dedicated high-intensity facilities - MEG II, Mu3e, as well as COMET and Mu2e. 
For flavour violating processes involving $\tau$-leptons, the situation is however much more promising, as searches at $\mu$TRISTAN vastly outperform future bounds on cLFV $\tau$-lepton decays, leading to a sensitivity reach that is several orders of magnitude larger.

\begin{figure}
    \centering
    \includegraphics[width=0.65\linewidth]{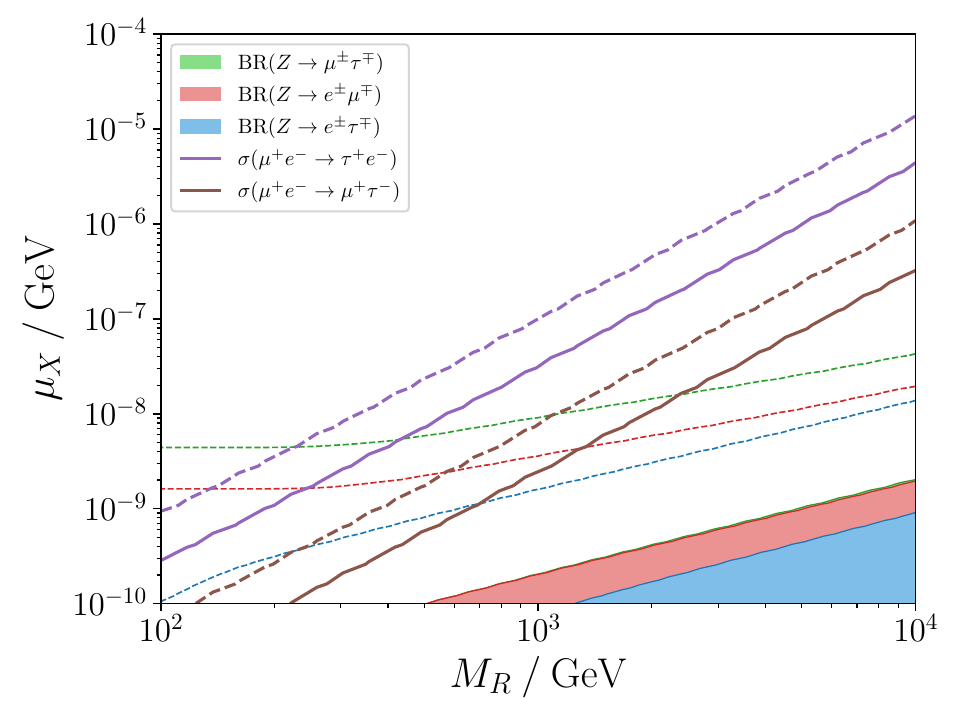}
    \caption{Prospects for ISS(3,3) cLFV searches depicted in the $(M_R-\mu_X)$ plane:
    current and future bounds of cLFV $Z$ decays (shaded regions, cf. Table~\ref{tab:cLFV_lep}) together with the projected sensitivity of $\mu$TRISTAN searches for $\mu^+e^-\to\tau^+e^-$ and $\mu^+e^-\to\mu^+\tau^-$ scatterings. 
    For the cross-sections, solid (dashed) lines correspond to $100\:\mathrm{ fb}^{-1}$ luminosity ($1\:\mathrm{ ab}^{-1}$).
    }
    \label{fig:contour_CI_Z}
\end{figure}

Interestingly, $\mu$TRISTAN is also very competitive with 
cLFV $Z$-boson decays - even in view of the most optimistic
expectations of a future Tera-$Z$ factory.
The contours for cLFV $Z$-boson decays, together  with $\mu$TRISTAN searches for $\mu^+e^-\to\tau^+e^-$ and $\mu^+e^-\to\mu^+\tau^-$ scatterings are shown in Figure~\ref{fig:contour_CI_Z}, for $100\:\mathrm{fb}^{-1}$ and $1\:\mathrm{ab}^{-1}$ luminosities.
As can be seen, the expected sensitivities of a future Tera-$Z$ factory are vastly outperformed by dedicated searches at $\mu$TRISTAN, highlighting the strong complementarity role of such a collider. 
This is further reinforced by recalling that searches at the $Z$-pole are in general plagued by irreducible systematics.

\section{Concluding remarks and outlook}
In this work we have addressed the sensitivity of an asymmetric lepton collider to cLFV signals, as those originating from the presence of heavy sterile states.
Focusing on $\mu$TRISTAN~\cite{Hamada:2022mua}, we have thus explored the prospects for observing cLFV lepton scatterings  $\mu^+ e^- \to \ell_\alpha^+ \ell_\beta^-$, studying two phenomenological frameworks: a minimal ad-hoc SM extension via two heavy neutral leptons, and a simple realisation of the Inverse Seesaw, the ISS(3,3).

We have carried out a full, detailed computation of the 
cross-sections for $\mu^+ e^- \to \ell_\alpha^+ \ell_\beta^-$ scattering, as well as other observables, such as rapidity distributions and forward-backward asymmetries. 
The asymmetries have been shown to offer valuable help in distinguishing 
the nature of dominant contributions - for instance loop- versus tree-level exchanges, and in the latter case provide important information on the mass of the mediators~\cite{Dev:2023nha}.
The full expressions for the observables are given in detail, and in a user-friendly way.

As a first approach to illustrate the potential of $\mu$TRISTAN
cLFV searches, we have considered a flavour-benchmark configuration for a minimal, ad-hoc SM extension via 2 HNLs (encoding their role in an enlarged leptonic mixing matrix).
Such a first, simple approach nevertheless clearly reveals that the reach of $\mu$TRISTAN might offer interesting, and clearly complementary probes in the search for cLFV at high-energies. Possible measurements of the rapidity distributions and $A_\text{FB}$ are further complementary and might offer valuable insight onto the underlying topology of the exchanges.
Furthermore, we have pointed out that such observables might also be sensitive to new sources of CP violation in the lepton sector.

These effects could be further explored with the possibility of beam polarisation (expected to be $\mathcal{O}(90\%)$ for $\mu^+$
and $\mathcal{O}(70\%)$ for $e^-$), which would allow to better identify the nature of the NP mediators.
Let us mention that here we have only considered fully leptonic processes - one could envisage addressing $\mu^+ e^-\to q\bar q$ production, as this would allow exploring possible connections to the most constraining low-energy cLFV transition (i.e. $\mu-e$ conversion, in view of the impressive future sensitivities), and possibly to leptonic meson decays. 

We have then considered a well-motivated model of neutrino mass generation, which constitutes a complete realisation featuring the essential building blocks present in the SM ad-hoc extension. Our study of the ISS supports our assessment of the role of $\mu$TRISTAN: not only is it complementary to low-energy cLFV searches, but it also offers much stronger probes for cLFV involving $\tau$-leptons (being remarkably sensitive to $Z$-penguin exchanges, even more than leptonic cLFV decays or cLFV $Z$ decays). 
Although we have not considered it here, one expects that such a probing power is also manifest for other low-energy seesaw realisation, as for instance the Linear Seesaw.

For $\tau-\ell$ flavour violation, 
$\mu$TRISTAN offers rich prospects for the observation of cLFV processes
- even in the absence of other signals ($Z$ decays and/or low-energy transitions). 
Strikingly, even for $5\times 10^{12}$ $Z$ bosons, due to an irreducible systematic misidentification of secondary leptons from the $\tau$ decays in $Z\to\tau\tau$ as primary leptons in $Z\to\ell\tau$, $\mu$TRISTAN is clearly more sensitive to $\mu-\tau$ flavour violation (via $Z$-penguin exchange) than FCC-ee.

\section*{Acknowledgements}
JK is supported by the Slovenian Research Agency under the research grant No. N1-0253 and in part by J1-3013. 
JK is grateful for the hospitality of the Theory group of the Laboratoire de physique de Clermont Auvergne where large parts of this work have been completed and JK is further grateful to Miha Nemev\v{s}ek for useful suggestions.
EP acknowledges financial support from the Swiss
National Science Foundation (SNF) under contract 200020 204428.
This project has received support from the European Union's Horizon 2020 research and innovation programme under the Marie Sk\l{}odowska-Curie grant agreement No.~860881 (HIDDe$\nu$ network) and from the IN2P3 (CNRS) Master Project, ``Hunting for Heavy Neutral Leptons'' (12-PH-0100).

\appendix

\section{Feynman rules for relevant vertices}\label{app:feynman}
The extended lepton sector will lead to modified charged and neutral current interactions. For the case of Majorana neutral fermions, the most relevant terms (for the purpose of the current study) are collected below. In particular, we include interactions with neutral and charged gauge bosons, Higgs and Goldstone bosons.
\begin{align}\label{eq:lagrangian:WGHZ}
& \mathcal{L}_{W^\pm}\, =\, -\frac{g_w}{\sqrt{2}} \, W^-_\mu \,
\sum_{\alpha=1}^{3} \sum_{j=1}^{3 + n_S} \mathcal{U}_{\alpha j} \bar \ell_\alpha 
\gamma^\mu P_L \nu_j \, + \, \text{H.c.}\,, \nonumber \\
& \mathcal{L}_{Z^0}^{\nu}\, = \,-\frac{g_w}{4 \cos \theta_w} \, Z_\mu \,
\sum_{i,j=1}^{3 + n_S} \bar \nu_i \gamma ^\mu \left(
P_L \mathcal{C}_{ij} - P_R \mathcal{C}_{ij}^* \right) \nu_j\,, \nonumber \\
& \mathcal{L}_{H^0}\, = \, -\frac{g_w}{4 M_W} \, H  \,
\sum_{i\ne j= 1}^{3 + n_S}    \bar \nu_i\,\left[\mathcal{C}_{ij}\,\left(
P_L m_i + P_R m_j \right) +\mathcal{C}_{ij}^\ast\left(
P_R m_i + P_L m_j \right) \right] \nu_j\ , \nonumber \\
& \mathcal{L}_{G^0}\, =\,\frac{i g_w}{4 M_W} \, G^0 \,
\sum_{i,j=1}^{3 + n_S}  \bar \nu_i \left[ \mathcal{C}_{ij}
\left(P_R m_j  - P_L m_i  \right) + \mathcal{C}_{ij}^\ast
\left(P_R m_i  - P_L m_j  \right)\right] \nu_j\,, \nonumber  \\
& \mathcal{L}_{G^\pm}\, =\, -\frac{g_w}{\sqrt{2} M_W} \, G^- \,
\sum_{\alpha=1}^{3}\sum_{j=1}^{3 + n_S} \mathcal{U}_{\alpha j}
\bar \ell_\alpha\left(
m_\alpha P_L - m_j P_R \right) \nu_j\, + \, \text{H.c.}\,.
\end{align}
We notice that the above interactions were cast in the physical lepton bases, with Greek indices denoting the flavour of the charged leptons, and $i,j=1 ... n_S$ the neutral fermion spectrum (with $n_S$ denoting the number of sterile states). Moreover, $\mathcal{C}_{ij} $ are defined as
\begin{equation}\label{eq:Cij}
    \mathcal{C}_{ij} = \sum_{\rho = 1}^3
  \mathcal{U}_{i\rho}^\dagger \,\mathcal{U}_{\rho j}^{\phantom{\dagger}}\:. 
\end{equation}

In what follows, and for completeness, we present the 
Feynman rules for the vertices which are used throughout our computation; these correspond to the modified Lagrangian terms which were given in Eq.~(\ref{eq:lagrangian:WGHZ}).
\begin{table}
\begin{tabular}{m{2.3cm}m{5.65cm}m{2.3cm}m{5.45cm}}
            \begin{tikzpicture}
    \begin{feynman}
    \vertex (a) at (0,0) {\(Z_\mu\)};
    \vertex (b) at (1,0);
    \vertex (c) at (2,1.){\(n_i\)};
    \vertex (d) at (2,-1.){\(n_j\)};
    \diagram* {
    (a) -- [boson] (b),
    (c) -- [momentum'=\( \)] (b),
    (b) -- [momentum'=\( \)] (d),
    };
    \end{feynman}
    \end{tikzpicture} 
    & $= \, -i \dfrac{g_w}{4 c_w} \gamma_\mu \left(C_{ij}^* \,P_L -C_{ij} \,P_R \right)$ 
    & & 
\\
\\
        \begin{tikzpicture}
    \begin{feynman}
    \vertex (a) at (0,0) {\(H\)};
    \vertex (b) at (1,0);
    \vertex (c) at (2,1.){\(n_i\)};
    \vertex (d) at (2,-1.){\(n_j\)};
    \diagram* {
    (a) -- [scalar] (b),
    (c) -- [momentum'=\( \)] (b),
    (b) -- [momentum'=\( \)] (d),
    };
    \end{feynman}
    \end{tikzpicture}
    & $=\, -i\dfrac{g_w}{4 M_W} \left[C_{ij} \left( m_i\, P_L + m_j \,P_R \right)
    + C_{ij}^* \left( m_i\, P_R + m_j \,P_L \right) \right]$
    \\
    \\
    \begin{tikzpicture}
    \begin{feynman}
    \vertex (a) at (0,0) {\(W_\mu^-\)};
    \vertex (b) at (1,0);
    \vertex (c) at (2.1,1.1){\(\nu_i\)};
    \vertex (d) at (2.1,-1.1){\(\ell^-\)};
    \vertex (e) at (1.9,0.55);
    \vertex (f) at (1.9,-0.55);
    \diagram* {
    (a) -- [boson] (b),
    (c) -- [ ] (b),
    (b) -- [fermion] (d),
    (f) -- [<-, quarter left, looseness=1] (e),
    };   
    \end{feynman}
    \end{tikzpicture}
    & $=\, -i\dfrac{g_w}{\sqrt{2}} \,\mathcal{U}_{\ell \nu}\, \gamma_\mu\, P_L$
    &
    %
    \begin{tikzpicture}
    \begin{feynman}
    \vertex (a) at (0,0) {\(W_\mu^+\)};
    \vertex (b) at (1,0);
    \vertex (c) at (2.1,1.1){\(\ell^+\)};
    \vertex (d) at (2.1,-1.1){\(\nu_i\)};
    \vertex (e) at (1.9,0.55);
    \vertex (f) at (1.9,-0.55);
    \diagram* {
    (a) -- [boson] (b),
    (c) -- [fermion] (b),
    (b) -- [ ] (d),
    (f) -- [<-, quarter left, looseness=1] (e),
    };
    \end{feynman}
    \end{tikzpicture}
    &
    $=\, -i\dfrac{g_w}{\sqrt{2}} \,\mathcal{U}_{\ell \nu}^*\, \gamma_\mu\, P_L$
    \\
    \\
    \begin{tikzpicture}
    \begin{feynman}
    \vertex (a) at (0,0) {\(G^-\)};
    \vertex (b) at (1,0);
    \vertex (c) at (2.1,1.1){\(\nu_i\)};
    \vertex (d) at (2.1,-1.1){\(\ell^-\)};
    \vertex (e) at (1.9,0.55);
    \vertex (f) at (1.9,-0.55);
    \diagram* {
    (a) -- [scalar] (b),
    (c) -- [ ] (b),
    (b) -- [fermion] (d),
    (f) -- [<-, quarter left, looseness=1] (e),
    };
    \end{feynman}
    \end{tikzpicture}  
    &
    $=\, i\dfrac{g_w}{\sqrt{2}M_W} \,\mathcal{U}_{\ell \nu} \,\left(m_\nu\, P_R - m_\ell\, P_L\right)$
&
    %
    \begin{tikzpicture}
    \begin{feynman}
    \vertex (a) at (0,0) {\(G^+\)};
    \vertex (b) at (1,0);
    \vertex (c) at (2.1,1.1){\(\ell^+\)};
    \vertex (d) at (2.1,-1.1){\(\nu_i\)};
    \vertex (e) at (1.9,0.55);
    \vertex (f) at (1.9,-0.55);
    \diagram* {
    (a) -- [scalar] (b),
    (c) -- [fermion] (b),
    (b) -- [ ] (d),
    (f) -- [<-, quarter left, looseness=1] (e),
    };
    \end{feynman}
    \end{tikzpicture}
    &
    $=\, i\dfrac{g_w}{\sqrt{2}M_W} \,\mathcal{U}_{\ell \nu}^* \,\left(m_\nu \,P_L - m_\ell \,P_R\right)$
    \\
    \\
        \begin{tikzpicture}
    \begin{feynman}
    \vertex (a) at (0,0) {\(W_\mu^-\)};
    \vertex (b) at (1,0);
    \vertex (c) at (2.1,1.1){\(\ell^-\)};
    \vertex (d) at (2.1,-1.1){\(\nu_i\)};
    \vertex (e) at (1.9,0.55);
    \vertex (f) at (1.9,-0.55);
    \diagram* {
    (a) -- [boson] (b),
    (c) -- [anti fermion] (b),
    (b) -- [ ] (d),
    (f) -- [<-, quarter left, looseness=1] (e),
    };
    \end{feynman}
    \end{tikzpicture}
    &
    $=\, -i\dfrac{g_w}{\sqrt{2}} \,\mathcal{U}_{\ell \nu}\, \gamma_\mu\, (\textcolor{red}{-P_R}) $
    &
        \begin{tikzpicture}
    \begin{feynman}
    \vertex (a) at (0,0) {\(W_\mu^+\)};
    \vertex (b) at (1,0);
    \vertex (c) at (2.1,1.1){\(\nu_i\)};
    \vertex (d) at (2.1,-1.1){\(\ell^+\)};
    \vertex (e) at (1.9,0.55);
    \vertex (f) at (1.9,-0.55);
    \diagram* {
    (a) -- [boson] (b),
    (c) -- [ ] (b),
    (b) -- [anti fermion] (d),
    (f) -- [<-, quarter left, looseness=1] (e),
    };
    \end{feynman}
    \end{tikzpicture}
    & $=\, -i\dfrac{g_w}{\sqrt{2}} \,\mathcal{U}_{\ell \nu}^*\, \gamma_\mu\, (\textcolor{red}{-P_R})  $
    %
    \\
    \\
    %
    \begin{tikzpicture}
    \begin{feynman}
    \vertex (a) at (0,0) {\(G^-\)};
    \vertex (b) at (1,0);
    \vertex (c) at (2.1,1.1){\(\ell^-\)};
    \vertex (d) at (2.1,-1.1){\(\nu_i\)};
    \vertex (e) at (1.9,0.55);
    \vertex (f) at (1.9,-0.55);
    \diagram* {
    (a) -- [scalar] (b),
    (c) -- [anti fermion] (b),
    (b) -- [ ] (d),
    (f) -- [<-, quarter left, looseness=1] (e),
    };
    \end{feynman}
    \end{tikzpicture}
    &
    $=\, i\dfrac{g_w}{\sqrt{2}M_W} \,\mathcal{U}_{\ell \nu} \,\left(m_\nu \,P_R - m_\ell \,P_L \right)$
    &
    \begin{tikzpicture}
    \begin{feynman}
    \vertex (a) at (0,0) {\(G^+\)};
    \vertex (b) at (1,0);
    \vertex (c) at (2.1,1.1){\(\nu_i\)};
    \vertex (d) at (2.1,-1.1){\(\ell^+\)};
    \vertex (e) at (1.9,0.55);
    \vertex (f) at (1.9,-0.55);
    \diagram* {
    (a) -- [scalar] (b),
    (c) -- [ ] (b),
    (b) -- [anti fermion] (d),
    (f) -- [<-, quarter left, looseness=1] (e),
    };
    \end{feynman}
    \end{tikzpicture}  
    &
    $=\, i\dfrac{g_w}{\sqrt{2}M_W} \,\mathcal{U}_{\ell \nu}^* \,\left(m_\nu \, P_L - m_\ell \, P_R \right)$
\end{tabular}
\caption{
Feynman rules for $W$, $Z$ and Higgs interactions, together with Goldstone bosons in generic SM extensions via Majorana sterile fermions (such as the ad-hoc ``3+2" extension, or the ISS(3,3) here considered).
}\label{table:feynrulesMajorana}
\end{table}

Notice that for the case of the $Z$ and Higgs vertices, the arrows denote the momentum flow. Since all physical neutral leptons are assumed to be of Majorana nature, diagrams including at least one $Z n_i n_j$ or $H n_i n_j$ vertex must thus be symmetrised (factor 2).
In the case of the $W$ interactions, the curved arrow represents an arbitrary (fermion) flow, see~\cite{Denner:1992vza} for more details.

\section{Computation of box-diagram contributions: comments 
and detailed discussion
}
\label{app:boxes}
While the computation of the penguin diagrams can be performed in Feynman 't-Hooft gauge in a fairly straightforward manner, the box diagrams require significantly more effort and attention.
First of all we notice that diagrams (3) and (4) of Figure~\ref{fig:boxes} contain LNV interactions and thus one needs to use the fermion-number violating Feynman rules of~\cite{Denner:1992vza}, which we summarised in Table~\ref{table:feynrulesMajorana} of Appendix~\ref{app:feynman}.
This leads, prior to integration, to the following amplitudes:
\begin{eqnarray}
    \mathcal M_1^{\alpha\beta} &=& \int\frac{d^D k}{(2\pi)^D}\,\bar v_\mu(p_1)\left[\left(\frac{-i g}{\sqrt{2}}\right)\mathcal U_{\mu i}\, \gamma_\alpha P_L\, i\frac{(\slashed{k}  + m_i)}{k^2 - m_i^2}\left(\frac{-i g}{\sqrt{2}}\right)\mathcal U_{e i}\, \gamma_\mu P_L\right]u_e(p_2)\nonumber\\
     &\phantom{=}&\times \,\bar u_\beta (p_4)\left[\left(\frac{-i g}{\sqrt{2}}\right)\mathcal U_{\beta j}^\ast\, \gamma_\nu P_L i \frac{(\slashed{k} + \slashed{p_1} - \slashed{p_3} + m_j)}{(k + p_1 - p_3)^2 - m_j^2}\left(\frac{-i g}{\sqrt{2}}\right)\mathcal U_{\alpha j}^\ast\, \gamma_\beta P_L\right]v_\alpha(p_3)\nonumber\\
     &\phantom{=}&\times\,\frac{(-i g^{\alpha\beta})}{(k+p_1)^2 - m_W^2}\frac{(-i g^{\mu\nu})}{(k -p_2)^2 - m_W^2}\,,\\
     \mathcal M_2^{\alpha\beta} &=& \int\frac{d^D k}{(2\pi)^D}\,\bar v_\mu(p_1)\left[\left(\frac{-i g}{\sqrt{2}}\right)\mathcal U_{\mu i}\, \gamma_\alpha P_L\, i\frac{(\slashed{k} - \slashed{p_1}  + m_i)}{(k - p_1)^2 - m_i^2}\left(\frac{-i g}{\sqrt{2}}\right)\mathcal U_{\alpha i}^\ast\, \gamma_\mu P_L\right]v_\alpha(p_3)\nonumber\\
     &\phantom{=}&\times \,\bar u_\beta (p_4)\left[\left(\frac{-i g}{\sqrt{2}}\right)\mathcal U_{\beta j}\, \gamma_\nu P_L i \frac{(\slashed{k} + \slashed{p_2} + m_j)}{(k + p_2)^2 - m_j^2}\left(\frac{-i g}{\sqrt{2}}\right)\mathcal U_{e j}^\ast\, \gamma_\beta P_L\right]u_e(p_2)\nonumber\\
     &\phantom{=}&\times\,\frac{(-i g^{\alpha\beta})}{k^2 - m_W^2}\frac{(-i g^{\mu\nu})}{(k - p_1 + p_3)^2 - m_W^2}\,,\\
    \mathcal M_3^{\alpha\beta} &=& \int\frac{d^D k}{(2\pi)^D}\,\bar v_\mu(p_1)\left[\left(\frac{-i g}{\sqrt{2}}\right)\mathcal U_{\mu i}\, \gamma_\alpha P_L\, i\frac{(\slashed{k} - \slashed{p_1} + m_i)}{(k - p_1)^2 - m_i^2}\left(\frac{-i g}{\sqrt{2}}\right)\mathcal U_{\beta i}\, \gamma_\mu (\textcolor{red}{-P_R})\right]\textcolor{teal}{\pmb{v_\beta}} (p_4)\nonumber\\
     &\phantom{=}&\times\, \textcolor{teal}{\pmb{\bar u_\alpha}} (p_3)\left[\left(\frac{-i g}{\sqrt{2}}\right)\mathcal U_{\alpha j}^\ast\, \gamma_\nu (\textcolor{red}{-P_R}) i \frac{(\slashed{k} + \slashed{p_2} + m_j)}{(k + p_2)^2 - m_j^2}\left(\frac{-i g}{\sqrt{2}}\right)\mathcal U_{e j}^\ast\, \gamma_\beta P_L\right]u_e(p_2)\nonumber\\
     &\phantom{=}&\times\,\frac{(-i g^{\alpha\beta})}{k^2 - m_W^2}\frac{(-i g^{\mu\nu})}{(k + p_2 - p_3)^2 - m_W^2}\,,\\
     \mathcal M_4^{\alpha\beta} &=& \int\frac{d^D k}{(2\pi)^D}\, \bar v_\mu(p_1)\left[\left(\frac{-i g}{\sqrt{2}}\right)\mathcal U_{\mu i}\, \gamma_\alpha P_L\, i\frac{(\slashed{k} - \slashed{p_1} + \slashed{p_3} + m_i)}{(k - p_1 + p_3)^2 - m_i^2}\left(\frac{-i g}{\sqrt{2}}\right)\mathcal U_{\beta i}\, \gamma_\mu (\textcolor{red}{-P_R})\right]\textcolor{teal}{\pmb{v_\beta}} (p_4)\nonumber\\
     &\phantom{=}&\times \,\textcolor{teal}{\pmb{\bar u_\alpha}}(p_3)\left[\left(\frac{-i g}{\sqrt{2}}\right)\mathcal U_{\alpha j}^\ast\, \gamma_\beta (\textcolor{red}{-P_R}) i \frac{(-\slashed{k} + m_j)}{k^2 - m_j^2}\left(\frac{-i g}{\sqrt{2}}\right)\mathcal U_{e j}^\ast\, \gamma_\nu P_L\right]u_e(p_2)\nonumber\\
     &\phantom{=}&\times\,\frac{(-i g^{\alpha\beta})}{(k + p_3)^2 - m_W^2}\frac{(-i g^{\mu\nu})}{(k + p_2)^2 - m_W^2}\,,
\end{eqnarray}
in which we colour-highlight the flip $P_L\to - P_R$ in the fermion-number violating $W$-vertices (cf. Appendix~\ref{app:feynman}) as well as the spinor-flip $u\leftrightarrow v$ (see \cite{Denner:1992vza} for details).
Moreover, one has to add the corresponding Goldstone diagrams to obtain physical amplitudes;
we stress here that the corresponding Goldstone vertices do not feature the $P_L\leftrightarrow - P_R$ flip as shown in Table~\ref{table:feynrulesMajorana}.

\bigskip
After extensive
algebraic transformations (or after using software like \texttt{package-X}~\cite{Patel:2015tea} or \texttt{FeynCalc}~\cite{Shtabovenko:2016sxi,Shtabovenko:2020gxv,Shtabovenko:2023idz} as done here), one encounters ``unusual'' Dirac structures in the integrated and simplified amplitudes (in $D=4-2\varepsilon$ dimensions), which can be reduced to the chiral Dirac basis with the help of the identities derived in Appendix~\ref{app:diracology}.
Here, we just summarise the results which we obtain in the limit of massless external fermions.

\bigskip
The amplitudes of box-diagrams (1) and (2) admit, in addition to $[\bar v \gamma_\mu P_L v][\bar u \gamma^\mu P_L u]$, the following ``unusual'' Dirac structures (with the short-hand notation: $u_i/v_{i} \equiv u/v(p_i)$)
\begin{equation}
    [\bar v_1 \slashed{p_3} P_L u_2]\,[\bar u_4 \slashed{p_2}P_L v_3] = - [\bar v_1 \slashed{p_3} P_L u_2]\,[\bar u_4 \slashed{p_1}P_L v_3]\,,
\end{equation}
and
\begin{equation}
    [\bar v_1 \slashed{p_2} P_L v_3]\,[\bar u_4 \slashed{p_3} P_L u_2] = - [\bar v_1 \slashed{p_2} P_L v_3]\,[\bar u_4 \slashed{p_1} P_L u_2]\,,
\end{equation}
in which we used on-shell conditions and the masslessness of the external fermions.
Since the box amplitudes are all finite, we can safely take the limit of $\varepsilon\to0$, that is taking the limit of $D=4$ spacetime dimensions.

Thus, setting external fermions to be massless and taking the limit $D=4$, the above Dirac structures can further be reduced to 
\begin{eqnarray}
    [\bar v_1 \slashed{p_3} P_L u_2]\,[\bar u_4 \slashed{p_1}P_L v_3] &=& g_{\alpha\beta}p_1^\alpha p_3^\beta [\bar v_1 \gamma_\mu P_L u_2]\,[\bar u_4 \gamma^\mu P_L v_3]\,,\label{eqn:idslsl1}\\
    {[\bar v_1 \slashed{p_2} P_L v_3]}\,{[\bar u_4 \slashed{p_1} P_L u_2]} &=&  g_{\alpha\beta}p_1^\alpha p_2^\beta[\bar v_1 \gamma_\mu P_L v_3]\,[\bar u_4 \gamma^\mu P_L u_2]\,.\label{eqn:idslsl2}
\end{eqnarray}

The remaining diagrams admit a plethora of further Dirac structures which can be reduced to a scalar combination
multiplied by a scalar product of momenta.
These are given by
\begin{eqnarray}
    [\bar v_1 \sigma_{\mu\alpha} P_R v_4]\,[\bar u_3 \sigma^{\mu\beta} P_L u_2] p_3^\alpha p_{1\beta} &=& -g_{\alpha\beta} p_3^{\alpha}p_1^{\beta} S_{RL}\,,\label{eqn:idsigsig1}\\{}
    [\bar v_1 \sigma_{\mu\alpha} P_R v_4]\,[\bar u_3 \sigma^{\mu\beta} P_L u_2] p_2^\alpha p_{1\beta} &=& g_{\alpha\beta} p_2^{\alpha}p_1^{\beta} S_{RL}\,,\label{eqn:idsigsig2}\\{}
    [\bar v_1 \sigma_{\mu\alpha} P_R v_4]\,[\bar u_3 \sigma^{\mu\beta} P_L u_2] p_4^\alpha p_{1\beta} &=& g_{\alpha\beta} (p_2^\alpha + p_3^{\alpha})p_1^{\beta} S_{RL}\,,\label{eqn:idsigsig3}\\{}
    [\bar v_1 \sigma_{\alpha\beta} P_R v_4]\,[\bar u_3 P_L u_2] p_2^\alpha p_{3}^\beta &=& i g_{\alpha\beta} (p_1^{\alpha} + p_4^{\alpha})p_1^{\beta} S_{RL}\,,\label{eqn:idsig1}\\{}
    [\bar v_1 \sigma_{\alpha\beta} P_R v_4]\,[\bar u_3 P_L u_2] p_1^\alpha p_{3}^\beta &=& -i g_{\alpha\beta} p_1^{\alpha}p_3^{\beta} S_{RL}\,,\label{eqn:idsig2}\\{}
    [\bar v_1 \sigma_{\alpha\beta} P_R v_4]\,[\bar u_3 P_L u_2] p_1^\alpha p_{2}^\beta &=& -i g_{\alpha\beta} p_1^{\alpha}p_2^{\beta} S_{RL}\,,\label{eqn:idsig3}
\end{eqnarray}
with $\sigma^{\mu\nu} = \dfrac{i}{2} [\gamma^{\mu}, \gamma^{\nu}]$ and brackets denote the commutator.
We have also introduced the abbreviation 
\begin{equation}
    S_{RL} = [\bar v_1 P_R v_4]\,[\bar u_3 P_L u_2]\,.\label{eqn:SRL}
\end{equation}
We stress again that these identities are only valid in the limit of massless external fermions~\footnote{Similar identities were obtained in~\cite{2405.02819}.}. The derivation of most of these identities require several non-trivial steps, which we outline in the following appendix.

\section{Notes on Diracology}
\label{app:diracology}
In the interest of reproducibility, and for the reader's convenience, we outline the main steps necessary to find the identities given in Eqs.~(\ref{eqn:idslsl1}-\ref{eqn:idsig3}).
First of all, let us note, since we work exclusively in the limit of massless external fermions, that all expressions are valid for $u$ and $v$ spinors, whose notation we unify as $w$ in the remainder of this section.
Furthermore, we reproduce here important identities of $\gamma$-matrices which are frequently used in the following.
This is the case of the so-called Gordon identity
\begin{eqnarray}
    \gamma^\mu\gamma^{\nu} &=&\frac{1}{2}\{\gamma^\mu, \gamma^\nu\} + \frac{1}{2}[\gamma^\mu, \gamma^\nu]\nonumber\\
    &=& g^{\mu\nu} - i \sigma^{\mu\nu} = g^{\mu\nu} + i \sigma^{\nu\mu}\,.\label{eqn:gordon}
\end{eqnarray}
A less common but nevertheless extremely useful identity is 
\begin{eqnarray}
    \gamma^{\mu}\gamma^\rho\gamma^\nu = g^{\mu\rho}\gamma^\nu - g^{\mu\nu} \gamma^\rho + g^{\rho\nu}\gamma^{\mu} - i \varepsilon^{\mu\rho\nu\alpha}\gamma_\alpha \gamma_5\,,
\end{eqnarray}
with the Levi-Civita tensor $\varepsilon_{0123}=1$ and $\gamma_5 = \frac{i}{4\!}\varepsilon_{\alpha\beta\mu\nu}\gamma^{\alpha}\gamma^\beta\gamma^\mu\gamma^\nu$.
Based on this identity one can show several useful identities of four-fermion current products, first derived in~\cite{Sirlin:1981pi}.
We reproduce here its main results (Eqs.~(1)-(4) of~\cite{Sirlin:1981pi})
\begin{eqnarray}
    [\bar w_3\gamma^\mu\gamma^\rho\gamma^\nu P_{L,R} w_1]\,[\bar w_4 \gamma_\mu \gamma^\tau\gamma_\nu P_{L,R} w_2] &=& 4 g^{\rho \tau}[\bar w_3 \gamma^\lambda P_{L,R} w_1]\,[\bar w_4 \gamma_\lambda P_{L,R} w_2]\,,\label{eqn:Sirlin1}\\{}
    [\bar w_3\gamma^\mu\gamma^\rho\gamma^\nu P_{L,R} w_1]\,[\bar w_4 \gamma_\mu \gamma^\tau\gamma_\nu P_{R,L} w_2] &=& 4 [\bar w_3 \gamma^\tau P_{L,R} w_1]\,[\bar w_4 \gamma^\rho P_{R,L} w_2]\,,\label{eqn:Sirlin2}\\{}
    [\bar w_3\gamma^\nu\gamma^\rho\gamma^\mu P_{L,R} w_1]\,[\bar w_4 \gamma_\mu \gamma^\tau\gamma_\nu P_{L,R} w_2] &=& 4 [\bar w_3 \gamma^\tau P_{L,R} w_1]\,[\bar w_4 \gamma^\rho P_{L,R} w_2]\,,\label{eqn:Sirlin3}\\{}
    [\bar w_3\gamma^\nu\gamma^\rho\gamma^\mu P_{L,R} w_1]\,[\bar w_4 \gamma_\mu \gamma^\tau\gamma_\nu P_{R,L} w_2] &=& 4 g^{\rho \tau}[\bar w_3 \gamma^\lambda P_{L,R} w_1]\,[\bar w_4 \gamma_\lambda P_{R,L} w_2]\,.\label{eqn:Sirlin4}
\end{eqnarray}
Another very useful identity appears in a footnote of~\cite{Sirlin:1981pi} and is given by
\begin{eqnarray}
    [\bar w_3 \gamma^\mu\gamma^\nu P_{L,R} w_1]\,[\bar w_4 \gamma_{\mu}\gamma_\nu P_{R,L} w_2] = 4 [\bar w_3 P_{L,R} w_1]\,[\bar w_4 P_{R,L} w_2]\,.\label{eqn:Sirlinf}
\end{eqnarray}
So far, the only condition used is that the dimensionality of spacetime is $D=4$, meaning that these identities are valid independent of external fermion masses.

\bigskip
Let us then begin with the first two identities in Eqs.~\eqref{eqn:idslsl1} and ~\eqref{eqn:idslsl2}:
one first makes use of the identity in Eq.~\eqref{eqn:Sirlin3} to obtain
\begin{equation}
    p_2^\alpha p_1^\beta [\bar w_1 \gamma_\alpha P_L w_3][\bar w_4 \gamma_\beta P_L w_2] = \frac{p_2^\alpha p_1^\beta}{4}[\bar w_1 \gamma^\nu\gamma_\beta\gamma^\mu P_L w_3][\bar w_4 \gamma_\mu\gamma_\alpha \gamma_\nu P_L w_2]\,.
\end{equation}
Although appearing significantly more complicated
at first glance, we can anti-commute $\gamma_\beta$ to the left-most and $\gamma_\alpha$ to the right-most position and apply the (massless) equations of motion, $\bar w_1 \slashed{p_1} = 0$ and $\slashed{p_2}w_2 = 0$, which lead us directly to 
\begin{eqnarray}
    \frac{p_2^\alpha p_1^\beta}{4}[\bar w_1 (-\gamma_\beta\gamma^{\nu} + 2 g_{\beta}^{\phantom{\beta}\nu})\gamma^\mu P_L w_3]\,[\bar w_4 \gamma_\mu(-\gamma_\nu \gamma_\alpha  + 2g_{\alpha\nu}) P_L w_2] = 
   \nonumber \\ 
=    \frac{p_2^\alpha p_1^\beta}{4}\cdot 2\cdot 2\cdot g_{\alpha\nu}g_\beta^{\phantom{\beta}\nu}[\bar w_1 \gamma^\mu P_L w_3]\,[\bar w_4 \gamma_\mu P_L w_2]\,.
\end{eqnarray}
The main focus here  is to notice that in the left Dirac bracket we had a ``slashed momentum'' (which we could only annihilate in the right Dirac bracket) and vice-versa. With the help of Eq.~\eqref{eqn:Sirlin3} one switches the externally contracted $\gamma$-matrices between Dirac brackets thus enabling these simplifications.

In the same spirit one can derive the identities shown in Eqs.~\eqref{eqn:idsig1} and~\eqref{eqn:idsig2}, albeit in a significantly more complicated way.
First of all, one notices in the derivation of Eqs.~(\ref{eqn:Sirlin1}-\ref{eqn:Sirlin4}), see~\cite{Sirlin:1981pi} for details, that one can insert an additional $\gamma$-matrix completely to the left or right in each of the brackets; after a straight-forward computation (closely following~\cite{Sirlin:1981pi}), one is led to
\begin{eqnarray}
    [\bar w_3\gamma^\alpha\gamma^\mu\gamma^\rho\gamma^\nu P_{L,R} w_1]\,[\bar w_4 \gamma_\alpha\gamma_\mu \gamma^\tau\gamma_\nu P_{L,R} w_2] &=& 4 g^{\rho \tau}[\bar w_3 \gamma^\alpha\gamma^\lambda P_{L,R} w_1]\,[\bar w_4 \gamma_\alpha\gamma_\lambda P_{L,R} w_2]\,,\label{eqn:2Sirlin1}\\{}
    [\bar w_3\gamma^\alpha\gamma^\mu\gamma^\rho\gamma^\nu P_{L,R} w_1]\,[\bar w_4 \gamma_\alpha\gamma_\mu \gamma^\tau\gamma_\nu P_{R,L} w_2] &=& 4 [\bar w_3 \gamma^\alpha\gamma^\tau P_{L,R} w_1]\,[\bar w_4 \gamma_\alpha\gamma^\rho P_{R,L} w_2]\,,\label{eqn:2Sirlin2}\\{}
    [\bar w_3\gamma^\alpha\gamma^\nu\gamma^\rho\gamma^\mu P_{L,R} w_1]\,[\bar w_4 \gamma_\alpha\gamma_\mu \gamma^\tau\gamma_\nu P_{L,R} w_2] &=& 4 [\bar w_3 \gamma^\alpha\gamma^\tau P_{L,R} w_1]\,[\bar w_4 \gamma_\alpha\gamma^\rho P_{L,R} w_2]\,,\label{eqn:2Sirlin3}\\{}
    [\bar w_3\gamma^\alpha\gamma^\nu\gamma^\rho\gamma^\mu P_{L,R} w_1]\,[\bar w_4 \gamma_\alpha\gamma_\mu \gamma^\tau\gamma_\nu P_{R,L} w_2] &=& 4 g^{\rho \tau}[\bar w_3 \gamma^\alpha\gamma^\lambda P_{L,R} w_1]\,[\bar w_4 \gamma_\alpha\gamma_\lambda P_{R,L} w_2]\,.\label{eqn:2Sirlin4}
\end{eqnarray}

\medskip
Starting from the left-hand side of Eq.~\eqref{eqn:idsigsig1}, we can first insert the Gordon identity and apply the equations of motion to obtain
\begin{eqnarray}
    A \equiv p_3^\alpha p_1^\beta[\bar w_1 \sigma_{\mu \alpha} P_R w_4]\,[\bar w_3 \sigma^{\mu}_{\phantom{\mu}\beta} P_L w_2] &=& -p_3^{\alpha} p_1^\beta \Big(g_{\alpha\beta} S_{RL} + [\bar w_1 \gamma^{\mu}\gamma_{\alpha} P_R w_4]\,[\bar w_3 \gamma_\mu\gamma_\beta P_L w_2]\Big)\,.\label{eqn:A}
\end{eqnarray}
In the next step, we use the identity in Eq.~\eqref{eqn:2Sirlin2} to rewrite the last second term of Eq.~\eqref{eqn:A}, giving us
\begin{equation}
    \tilde A\equiv p_3^\alpha p_1^\beta[\bar w_1 \gamma^{\mu}\gamma_{\alpha} P_R w_4]\,[\bar w_3 \gamma_\mu\gamma_\beta P_L w_2] = \frac{p_3^\alpha p_1^\beta}{4} [\bar w_1 \gamma^\lambda \gamma^\mu\gamma_\beta\gamma^\nu P_R w_4]\,[\bar w_3 \gamma_\lambda \gamma_\mu\gamma_\alpha\gamma_\nu P_L w_2]\,.
\end{equation}
Now one again  anti-commutes $\gamma_\alpha$ and $\gamma_\beta$ to the left-most slots and apply the equations of motion, leading to
\begin{equation}
    \tilde A = p_3^\alpha p_1^\beta \Big(8 g_{\alpha\beta}S_{RL} - 2 [\bar w_1 \gamma_\alpha\gamma^\nu P_R w_4]\,[\bar w_3 \gamma_\beta \gamma_\nu P_L w_2] \Big)\,.
\end{equation}
Although the gain may not seem apparent,  
note the different order of $\gamma$-matrices in the second bracket. 
At this point one 
re-inserts the Gordon identity (this time the third equality of Eq.\eqref{eqn:gordon}) which gives us back the original term $A$ as
\begin{equation}
    \tilde A = p_3^\alpha p_1^\beta \Big(2 g_{\alpha\beta}S_{RL}  + 2 A\Big)\,,
\end{equation}
such that finally we have 
\begin{eqnarray}
    A &=& -p_3^{\alpha} p_1^{\beta} \Big(g_{\alpha\beta} S_{RL} + 2 g_{\alpha\beta} S_{RL} + 2A\Big)\,,\\
    \Rightarrow A &=& -p_3^\alpha p_1^\beta g_{\alpha\beta} S_{RL}\,.
\end{eqnarray}
(Leading to the obtention of this result one can also quickly show Eqs.\eqref{eqn:idsig2} by inserting once again the Gordon identity.)

The identity in Eq.\eqref{eqn:idsigsig2} is probably the most cumbersome and the least evident:
firstly, one notices that in deriving Eq.\eqref{eqn:idsigsig1} both momenta had to be annihilated to the left of the brackets, while now we have one to the right and one to the left.
One possible way to overcome this is to use the momentum conservation $p_1 + p_2 = p_3 + p_4$ to rewrite $p_2 = p_3 + p_4 - p_1$, so that one quickly shows that
\begin{equation}
    [\bar w_1 \sigma_{\mu\alpha} P_R w_4]\,[\bar w_3 \sigma^{\mu\beta} P_L w_2] p_1^\alpha p_{1\beta} = 0\,,
\end{equation}
again by inserting, 
the Gordon identity and using the fact that $\slashed p_1 \slashed p_1 = p_1^2 = 0$.
Similarly, one can derive and use Eq.~\eqref{eqn:idsigsig3},
which after insertion of the Gordon identity and $p_4 = p_1 + p_2 - p_3$ reduces to 
\begin{equation}
    [\bar w_1  P_R w_4]\,[\bar w_3 \sigma_{\alpha\beta} P_L w_2] p_4^\alpha p_{1\beta} \,=\, i g_{\alpha\beta} (p_2^{\alpha} + p_3^{\alpha})p_1^{\beta} S_{RL}\,.
\end{equation}
Adding this term to our result of Eq.~\eqref{eqn:idsigsig1}, we obtain Eq.~\eqref{eqn:idsigsig2}.
The remaining identities of Eqs.~\eqref{eqn:idsig2} and~\eqref{eqn:idsig3} can be derived by once again inserting the Gordon identity and applying the equations of motion.

Following the derivations in this appendix, more identities can be obtained in the limit of massless external fermions.
For massive fermions the expressions become significantly more complicated and are left for future work should external fermion masses become relevant.

{\small
\bibliographystyle{JHEP}
\bibliography{Mutristan}
}

\end{document}